\documentclass[times,twocolumn,final,longtitle]{elsarticle}

\usepackage{medima}
\usepackage{framed,multirow}
\usepackage{booktabs} 
\usepackage{amssymb}
\usepackage{latexsym}
\usepackage{hyperref}
\usepackage{amsmath}
\usepackage{enumitem}
\usepackage{multirow}
\usepackage{makecell}
\usepackage[german, english]{babel}
\usepackage{url}
\usepackage{xcolor}

\definecolor{newcolor}{rgb}{.8,.349,.1}

\journal{Medical Image Analysis}

\begin{document}

\verso{Jun Lyu \textit{et~al.}}

\begin{frontmatter}

\title{The state-of-the-art in Cardiac MRI Reconstruction: Results of the CMRxRecon Challenge in MICCAI 2023}

\author[1]{Jun \snm{Lyu}\fnref{fn1}}
\fntext[fn1]{These authors contribute equally.}
\author[2]{Chen \snm{Qin}\fnref{fn1}}
\author[3]{Shuo \snm{Wang}\fnref{fn1}}
\author[4,5]{Fanwen \snm{Wang}\fnref{fn1}}
\author[6]{Yan \snm{Li}}
\author[7,4]{Zi \snm{Wang}}
\author[7]{Kunyuan \snm{Guo}}
\author[8]{Cheng \snm{Ouyang}}
\author[4,5]{Michael \snm{T\"{a}nzer}}
\author[9]{Meng \snm{Liu}}
\author[9]{Longyu \snm{Sun}}
\author[9]{Mengting \snm{Sun}}
\author[9]{Qin \snm{Li}}
\author[10]{Zhang \snm{Shi}}
\author[11]{Sha \snm{Hua}}
\author[12]{Hao \snm{Li}}
\author[12]{Zhensen \snm{Chen}}
\author[2]{Zhenlin \snm{Zhang}}
\author[13]{Bingyu \snm{Xin}}
\author[13]{Dimitris N. \snm{Metaxas}}
\author[14]{George \snm{Yiasemis}}
\author[14]{Jonas \snm{Teuwen}}
\author[15]{Liping \snm{Zhang}}
\author[15]{Weitian \snm{Chen}}
\author[16]{Yidong \snm{Zhao}}
\author[16]{Qian \snm{Tao}}
\author[17]{Yanwei \snm{Pang}}
\author[17]{Xiaohan \snm{Liu}}
\author[18]{Artem \snm{Razumov}}
\author[18,32]{Dmitry V. \snm{Dylov}}
\author[19]{Quan \snm{Dou}}
\author[19]{Kang \snm{Yan}}
\author[20]{Yuyang \snm{Xue}}
\author[20]{Yuning \snm{Du}}
\author[21]{Julia \snm{Dietlmeier}}
\author[22]{Carles \snm{Garcia-Cabrera}}
\author[23]{Ziad \snm{Al-Haj Hemidi}}
\author[24]{Nora \snm{Vogt}}
\author[25]{Ziqiang \snm{Xu}}
\author[26]{Yajing \snm{Zhang}}
\author[27]{Ying-Hua \snm{Chu}}
\author[28]{Weibo \snm{Chen}}
\author[8]{Wenjia \snm{Bai}}
\author[29]{Xiahai \snm{Zhuang}}
\author[30]{Jing \snm{Qin}}
\author[31]{Lianmin \snm{Wu}\corref{cor1}}
\cortext[cor1]{Corresponding author.}
\ead{wlmssmu@126.com }
\author[4,5]{Guang \snm{Yang}\corref{cor1}}
\ead{g.yang@imperial.ac.uk}
\author[7]{Xiaobo \snm{Qu}\corref{cor1}}
\ead{quxiaobo@xmu.edu.cn}
\author[9,12]{He \snm{Wang}\corref{cor1}}
\ead{hewang@fudan.edu.cn}
\author[9]{Chengyan \snm{Wang}\corref{cor1}}\ead{wangcy@fudan.edu.cn}

\address[1]{Psychiatry Neuroimaging Laboratory, Brigham and Women’s Hospital, Harvard Medical School, 399 Revolution Drive, Boston, 02215, MA, United States}
\address[2]{Department of Electrical and Electronic Engineering \& I-X, Imperial College London, United Kingdom}
\address[3]{Digital Medical Research Center, School of Basic Medical Sciences, Fudan University, Shanghai, China}
\address[4]{Department of Bioengineering \& I-X, Imperial College London, London W12 7SL, UK;}
\address[5]{Cardiovascular Magnetic Resonance Unit, Royal Brompton Hospital, Guy’s and St Thomas’ NHS Foundation Trust, London SW3 6NP, UK}
\address[6]{Department of Radiology, Ruijin Hospital, Shanghai Jiao Tong University School of Medicine, Shanghai, China}
\address[7]{Department of Electronic Science, Fujian Provincial Key Laboratory of Plasma and Magnetic Resonance, National Institute for Data Science in Health and Medicine, Institute of Artificial Intelligence, Xiamen University, Xiamen 361102, China }
\address[8]{Department of Computing \& Department of Brain Sciences, Imperial College London, London SW7 2AZ, UK}
\address[9]{Human Phenome Institute, Fudan University, 825 Zhangheng Road, Pudong New District, Shanghai, 201203, China}
\address[10]{Department of Radiology, Zhongshan Hospital, Fudan University, Shanghai, China}
\address[11]{Department of Cardiovascular Medicine, Ruijin Hospital Lu Wan Branch, Shanghai Jiao Tong University School of Medicine, Shanghai, China}
\address[12]{Institute of Science and Technology for Brain-Inspired Intelligence, Fudan University, Shanghai, China, 200433}
\address[13]{Department of Computer Science, Rutgers University, Piscataway, NJ 08854, USA}
\address[14]{AI for Oncology, Netherlands Cancer Institute, Plesmanlaan 121, 1066 CX, Amsterdam, Netherlands}
\address[15]{CUHK Lab of AI in Radiology (CLAIR), Department of Imaging and Interventional Radiology, The Chinese University of Hong Kong, China}
\address[16]{Department of Imaging Physics, Delft University of Technology
Lorentzweg 1, 2628CN, Delft, Netherlands}
\address[17]{TJK-BIIT Lab, School of Electrical and Information Engineering, Tianjin University, Tianjin 300072, China}
\address[18]{Skolkovo Institute Of Science And Technology, Center for Artificial Intelligence Technology, 30/1 Bolshoy blvd., 121205 Moscow, Russia}
\address[19]{Department of Biomedical Engineering, University of Virginia,
415 Lane Rd., Charlottesville, VA 22903, United States}
\address[20]{Institute for Imaging, Data and Communications, University of Edinburgh, EH9 3FG, UK}
\address[21]{Insight SFI Research Centre for Data Analytics, Dublin City University, Glasnevin Dublin 9 Ireland}
\address[22]{ML-Labs SFI Centre for Research Training in Machine Learning, Dublin City University, Glasnevin Dublin 9 Ireland}
\address[23]{Institute of Medical Informatics, Universität zu Lübeck, Ratzeburger Alle 160,  23562 Lübeck, Germany}
\address[24]{IADI, INSERM U1254, Bâtiment Recherche, CHRU de Nancy Brabois, Rue du Morvan, 54511 Vandoeuvre-lès-Nancy, France}
\address[25]{School of Health Science and Engineering, University of Shanghai for Science and Technology, Shanghai, China}
\address[26]{MR Business Unit, Philips Healthcare Suzhou, China}
\address[27]{Siemens Healthineers Ltd., China}
\address[28]{Philips Healthcare, Shanghai, China}
\address[29]{School of Data Science, Fudan University, Shanghai, China}
\address[30]{School of Nursing, The Hong Kong Polytechnic University, Hong Kong, China}
\address[31]{Department of Radiology, Ren Ji Hospital, School of Medicine, Shanghai Jiao Tong University, Shanghai, 200127, China}
\address[32]{Artificial Intelligence Research Institute, 32/1 Kutuzovsky pr., Moscow, 121170, Russia}

\communicated{}

\begin{abstract}
Cardiac magnetic resonance imaging (MRI) provides detailed and quantitative evaluation of the heart's structure, function, and tissue characteristics with high-resolution spatial-temporal imaging. However, its slow imaging speed and motion artifacts are notable limitations. Undersampling reconstruction, especially data-driven algorithms, has emerged as a promising solution to accelerate scans and enhance imaging performance using highly under-sampled data. Nevertheless, the scarcity of publicly available cardiac k-space datasets and evaluation platform hinder the development of data-driven reconstruction algorithms. To address this issue, we organized the Cardiac MRI Reconstruction Challenge (CMRxRecon) in 2023, in collaboration with the 26th International Conference on Medical Image Computing and Computer-Assisted Intervention (MICCAI). CMRxRecon presented an extensive k-space dataset comprising cine and mapping raw data, accompanied by detailed annotations of cardiac anatomical structures. With overwhelming participation, the challenge attracted more than 285 teams and over 600 participants. Among them, 22 teams successfully submitted Docker containers for the testing phase, with 7 teams submitted for both cine and mapping tasks. All teams use deep learning based approaches, indicating that deep learning has predominately become a promising solution for the problem. The first-place winner of both tasks utilizes the E2E-VarNet architecture as backbones. In contrast, U-Net is still the most popular backbone for both multi-coil and single-coil reconstructions. This paper provides a comprehensive overview of the challenge design, presents a summary of the submitted results, reviews the employed methods, and offers an in-depth discussion that aims to inspire future advancements in cardiac MRI reconstruction models. The summary emphasizes the effective strategies observed in Cardiac MRI reconstruction, including backbone architecture, loss function, pre-processing techniques, physical modeling, and model complexity, thereby providing valuable insights for further developments in this field.
\end{abstract}


\end{frontmatter}


\section{Introduction}
\label{int}
\subsection{Background}
Cardiac magnetic resonance imaging (MRI) has emerged as a crucial imaging technique for non-invasive diagnosis in clinical cardiology, due to its advantages in quantitative assessment of cardiac morphology and myocardial tissue characteristics ~\citep{eyre2022simultaneous}. 

Cardiac cine (dynamic image sequence) offers a fine spatial and temporal resolution of the heart throughout the cardiac cycle. As the most common technique in cardiac MRI, cine can provide cardiac function measurements, e.g., cardiac output and ejection fraction~\citep{menchon2019reconstruction}.
In recent years, multi-contrast imaging methods, notably T1 mapping, and T2 mapping, have gained prominence in clinical applications~\citep{qi2021synergistic}. These mapping techniques exhibit high sensitivity in detecting lesions, allowing for the quantitative assessment of myocardial fibrosis, hemorrhage, and edema~\citep{jerosch2022cardiac}. In contrast to conventional methods, these techniques offer a direct measurement of the myocardial tissue's T1 and T2 attenuation values. These quantitative parameters can be utilized more effectively for the detection of diffuse lesions and provide a better benchmark for comparing measurements across multiple centers.
Although cardiac MRI offers numerous advantages, its main challenge lies in the slow scan speed it entails. Cine imaging captures multiple phases using segmented acquisition spread over different heartbeats, while mapping techniques require sampling multiple frames across the magnetization recovery to estimate T1 and T2 values. To achieve comprehensive coverage of the heart, repeated acquisitions from different orientations result in extended imaging time. This slow speed not only compromises the image quality due to the accumulation of imaging artifacts caused by patient movement but also exacerbates patient discomfort during the scanning process. 
To expedite cardiac MRI, the current accelerated solution involves compressed sensing (CS)\citep{donoho2006compressed,lustig2008compressed}. Instead of sampling the entire MRI signal space (Fourier space, usually termed as \textit{k-space}), CS only acquires a subset of k-space and recovers these sub-Nyquist measurements using iterative reconstruction algorithms. 
However, CS-MRI reconstructions have limited acceleration factors in practical use, and the introduction of certain regularization terms can compromise the fidelity and clarity of the resulting images. Additionally, the CS algorithm often requires prolonged computation time due to its iterative nature~\citep{lustig2008compressed,uecker2014espirit}. 
Therefore, the accurate and robust reconstruction of multi-contrast cardiac images from highly undersampled k-space data remains an open problem. 

In recent years, data-driven methods\citep{lyu2022dudocaf, lyu2023multi, lyu2023adaptive, lv2021transfer, lv2021pic, lv2021gan, lv2020parallel, lv2018reconstruction} have reshaped the general practice of image reconstruction. In addition to innovative network designs, the performance of these algorithms largely depends on the size and quality of the training dataset. To this end, several large-scale challenges, such as fastMRI~\citep{zbontar2018fastmri} and MC-MRI~\citep{10.3389/fnins.2022.919186}, have been organized for fair evaluations of these reconstruction algorithms. 
To date, there have been no dedicated challenges or publicly available datasets specifically focused on cardiac MRI reconstruction, not to mention the absence of a comprehensive and fair evaluation benchmark for the development of deep learning-enabled cardiac MRI reconstructions.

\subsection{Challenges of Cardiac MRI Reconstruction}
Cardiac MRI reconstruction faces several prominent challenges, including: 
\begin{itemize}
    \item Variable heart rhythms: Patients may have variable heart rates, leading to variations in the duration of cardiac cycles. This variability can impact the synchronization of data acquisition, requiring adaptive reconstruction methods to handle different heart rates effectively.

    \item Motion artifacts: The heart is a dynamic organ that undergoes continuous motion during the cardiac cycle resulting from inconsistencies across different segments of the acquisition process. Similarly, cardiac motion, as well as respiratory motion, may induce blurring in the reconstructed images, particularly when the acquisition window encompasses phases of rapid movement ~\citep{ismail2022cardiac}.

    \item Complex anatomy: The intricate anatomy of the heart, including delicate structures and complex geometries, requires reconstruction algorithms capable of preserving spatial details and accurately representing cardiac structures. On the other hand, if the training samples do not include certain diseases present in the test set, it can lead to a significant performance drop in deep-learning reconstruction models.

    \item Limited temporal resolution: Cardiac MRI involves capturing images at different phases of the cardiac cycle. Limited temporal resolution can result in inadequate coverage of dynamic events within the heart, which needs to be improved by accelerating the imaging and reducing the acquisition window.

\item Integration of deep learning with conventional methods: Technical solutions are still needed to effectively integrate conventional parallel imaging with deep learning techniques to maximize their respective advantages.

    \item High demands on computational resources: The raw data acquired during cardiac MRI scans, often coupled with the need for real-time reconstruction in some cases, poses significant computational challenges. Efficient algorithms and powerful hardware are essential for timely reconstruction.

\end{itemize}

There have been continuous efforts to develop advanced image reconstruction techniques, from compressed sensing to deep learning approaches, to address these challenges and improve the overall quality and efficiency of cardiac MRI reconstruction. The outcome of these advancements will be expected to contribute to more accurate diagnoses and better patient outcomes in cardiac imaging.

\subsection{Limitation of Existing Datasets}
To date, NYU Langone Health has released the ``fastMRI'' dataset, containing the multi-coil brain, knee, and prostate MRI raw k-space data. Similarly, the Universities of Calgary and Campinas have provided the MRI community with the ``Calgary-Campinas`` dataset~\citep{souza2018open}, comprising multi-coil brain acquisitions. However, these datasets do not apply to the spatio-temporal scenario in cardiac imaging. To the best of our knowledge, previous available cardiac raw datasets mainly include OCMR ~\citep{chen2020ocmr} and Harvard CMR Dataverse ~\citep{DVN/CI3WB6_2020}. The former provides fully sampled cine data as well as prospectively undersampled data, while the latter also offers cine data with radial sampling trajectories, including 101 patients and 7 healthy volunteers. However, these datasets suffer from limitations in a lack of anatomical views, imaging contrasts, and size of the dataset.
In comparison, our CMRxRecon dataset aims to provide a larger data size (300 subjects), more imaging contrasts (cine, T1 mapping, T2 mapping), and more anatomical views (short-axis, 2/3/4-chambers). Additionally, the CMRxRecon dataset is currently the only publicly released cardiac MRI reconstruction dataset associated with an open challenge.

\subsection{CMRxRecon Challenge}
CMRxRecon challenge is jointly organized by 10 institutions: Fudan University, Imperial College London, Hong Kong Polytechnic University, Xiamen University, the University of Texas, Shanghai Polytechnic University, Shanghai Jiao Tong University, Fudan University Affiliated Zhongshan Hospital, and Siemens Healthineers. 
It is a one-time event with fixed submission deadline. The CMRxRecon challenge aims to establish a platform for fast cardiac MRI reconstruction and provide a benchmark dataset that enables the broad research community to promote advances in this area of research, which includes two independent tasks:
\begin{itemize}
    \item Accelerated cine reconstruction
\end{itemize}
The aim of task 1 is to accelerate cine imaging from under-sampling data and address the image degradation problem caused by k-space under-samping. 
\begin{itemize}
    \item Accelerated T1 \& T2 mapping
\end{itemize}
The aim of task 2 is to improve the T1 and T2 mapping from under-sampling data and address the image degradation problem caused by k-space under-samping.

\subsection{Challenge Rules}
Each team can choose to participate one of them or both. The  top 3 winners in each task are invited to give oral presentations during the 26th International Conference on Medical Image Computing and Computer Assisted Intervention (MICCAI). The recommended authors of the top 3 winner teams are invited to contribute to the challenge summary paper. In addition, monetary awards are provided for the top 3 winners of each task. The prize pool is exclusively sponsored by Siemens Healthineers. 
Members of the organizers' institutes can participate but are not eligible for awards. Participating teams can publish their own results separately after the embargo time (three months after the announcement of the final results).

\subsection{Contributions}
The contributions of the ``CMRxRecon'' challenge include but are not limited to the following aspects:
\begin{itemize}
    \item Open dataset: CMRxRecon is the first cardiac MRI reconstruction challenge that provides an open dataset consisting of multi-contrast, multi-view, and multi-coil raw k-space data from 300 subjects with complete cardiac segmentation labels. This rich dataset holds crucial value for the development of deep learning algorithms. 
    \item Evaluation platform: Our challenge provides a benchmarking platform that enables timely evaluation of reconstruction results. Researchers can conveniently compare different algorithms using the same data and the same assessment metrics, thereby expediting their research progress and facilitating future research in the cardiac MRI field. 
    \item Methodology summary: Through the challenge, we evaluate and compare different deep-learning-based reconstruction methods on the two tasks, providing a summary of experiences and a comparison of the strengths and weaknesses of methods in the cardiac MRI reconstruction community. The summary highlights the effective strategies for CMR reconstruction, regarding the backbone architecture, loss function, pre-processing, physical model and model complexity, providing insights for further development.

\end{itemize}

In summary, the goal of establishing the CMRxRecon challenge is to provide a benchmark dataset that enables the broad research community to participate in this important work of accelerated CMR imaging. Through training and validation on this dataset using private models, we look forward to continuous technological breakthroughs in the field of cardiac MRI reconstruction, which will also contribute to the translation of the latest techniques into clinical practice. 
\section{Related Work}

\subsection{Cardiac MRI Challenges}
Over the last decade, there have been several challenges that focused on cardiac MRI. The majority of them focus on the segmentation of anatomical structures of the heart, such as the left ventricle (LV), the myocardium (Myo), and the right ventricle (RV). For example, one of the earliest cardiac MRI segmentation challenges held by the Cardiac Atlas Project~\citep{suinesiaputra2014collaborative} required participants to segment the Myocardium on steady-state free precession (SSFP) cine in short-axis images. The Right Ventricle Segmentation Challenge~\citep{petitjean2015right} focused on the segmentation of the RV on cine. Subsequenly, further related challenges emerged, including the Automated Cardiac Diagnosis Challenge (ACDC)~\citep{bernard2018deep} and the Multi-Centre, Multi-Vendor and Multi-Disease Cardiac Image Segmentation Challenge (M\&Ms)~\citep{campello2021multi} , which both consider the segmention of LV, Myo and RV. The ACDC challenge is composed of 150 subjects divided into five subgroups (normal, myocardial infarction, dilated cardiomyopathy, hypertrophic cardiomyopathy, and abnormal right ventricle) and the M\&Ms challenge additionally contributes to the effort of building generalizable models by providing CMR data scanned in clinical centers from three different countries using four different magnetic resonance scanner vendors, with a follow-up challenge on further incorporating multi-view CMR data ~\citep{martin2023deep}. Beyond cine CMR, other CMR sequences have been explored for the segmentation tasks. For instance, the Multi-Sequence Cardiac MR Segmentation Challenge (MS-CMRSeg)~\citep{zhuang2022cardiac} proposed to segment the ventricles and Myo using three CMR sequences, i.e. late gadolinium enhancement (LGE), T2 and bSSFP. The Atria Segmentation Challenge~\citep{xiong2021global} as well as the Left Atrial and Scar Quantification \& Segmentation (LAScarQS) Challenge~\citep{li2022atrialjsqnet} focused on the segmentation of left atrium and scar on LGE CMR. Whole heart segmentation has also been performed on both CT and MRI through the Multi-Modality Whole Heart Segmentation (MM-WHS) challenge~\citep{zhuang2019evaluation}.

Apart from the cardiac segmentation challenges, further challenges were held to tackle other CMR tasks, such as LV statistical shape modelling~\citep{suinesiaputra2017statistical} and classification of pathology~\citep{bernard2018deep,lalande2022deep}. Cardiac motion has also been of great interest in the CMR community, and challenges such as on cardiac motion analysis~\citep{tobon2013benchmarking} and motion correction~\citep{pontre2016open} have also been held. One recent challenge (CMRxMotion)~\citep{wang2022extreme} has proposed to establish a public benchmark dataset to assess the effects of respiratory motion on CMR imaging quality and examine the robustness of automated segmentation models. Despite the numerous efforts to organize challenges and establish benchmarks for CMR analysis, no benchmarks for upstream CMR reconstruction applications have been proposed to date. Our effort in this work and challenge aims to establish a platform for fast CMR image reconstruction and provide a benchmark dataset with both dynamic cine CMR and quantitative T1/T2 mapping raw data, for promoting advances in this area of research.
\subsection{Deep Learning Methods for Cardiac MRI Reconstruction}
Deep learning (DL) approaches have gained great popularity for MRI reconstruction in recent years, due to their excellent capabilities in reconstructing high-quality MR images at fast reconstruction speed. Current deep learning methods for cardiac MRI reconstruction can generally be categorized into three types~\citep{qin2022artificial}: image post-processing approaches, model-driven unrolled methods, and k-space-based interpolation techniques.

MRI reconstruction via image post-processing techniques typically learns an end-to-end mapping between zero-filled under-sampled images and ground truth fully-sampled references. Commonly, U-net architectures are employed to reduce image artefacts~\citep{hauptmann2019real,kofler2019spatio,lyu2023adaptive}, where 3D convolutions or 2D convolutions on the spatio-temporal domain ($x-t$) are leveraged to exploit the temporal information of cardiac MRI sequences. However, one significant drawback of this type of approach is that they do not consider the physically acquired $k$-space raw data during the reconstruction process and thus cannot guarantee the consistency between the reconstructed images and the acquired signals. To improve on this, model-driven, unrolled approaches~\citep{aggarwal2018modl,hammernik2018learning,qin2018convolutional,schlemper2017deep,duan2019vs,wang2023odls} have been proposed to embed the conventional iterative compressed sensing (CS)-based methods into deep learning frameworks. Such methods learn the unrolled optimization inspired by CS-based approaches, where the reconstruction process is alternated between a learnable image de-aliasing step parameterized by neural networks and a data consistency step. These models can be structured either in a cascaded fashion~\citep{schlemper2017deep} or in a recurrent way~\citep{qin2018convolutional} to mimic the iterative nature of the optimisation-based approaches. This type of method has been shown to be able to achieve state-of-the-art performance in CMR reconstruction with high fidelity and good generalization capability~\citep{hammernik2021systematic}, due to the incorporation of the physically acquired raw data within the learning process. Lastly, k-space interpolation approaches recover the missing data directly in k-space. For instance, CNNs or implicit neural representations can be used in k-space to learn the interpolation of k-space data given the auto-calibrating signals or the under-sampled signals~\citep{akccakaya2019scan,huang2023neural,feng2022spatiotemporal}. These approaches are typically subject-specific and do not require training datasets, but will need separate training for each scan.

The recent advancement of DL in CMR reconstruction has mainly been developed based on the above three types of approaches while incorporating some more advanced DL techniques. For instance, transformers have been studied in the context of CMR reconstruction to exploit the spatio-temporal information within and across cardiac frames~\citep{lyu2023region, lyu2024stadnet}, and diffusion models have been investigated to leverage their generative power for recovering high-quality scans within the above three frameworks~\citep{chung2022score,xie2022measurement,gungor2023adaptive}. A further research direction is to exploit information from complementary domains or use complementary regularisation. For instance, regularisation on spatial frequency domain along with that on image domain has been proposed to reconstruct the cine CMR~\citep{qin2019k,qin2021complementary}, which has demonstrated better performance compared to single domain reconstruction. Similarly, there have also been works on joint k-space and image space reconstruction~\citep{wang2022dimension}, as well as reconstruction methods incorporating low-rank or sparse prior~\citep{huang2021deep,wang2024deepssl} or SmooThness regularisation on manifolds (SToRM) prior~\citep{biswas2019dynamic}. Additionally, motion compensation on cine CMR has also been considered within the model-driven unrolled framework, where the reconstruction and motion are jointly estimated during the process~\citep{seegoolam2019exploiting,pan2024unrolled}. For a more comprehensive review of the existing DL approaches for CMR reconstruction, please refer to~\citep{qin2022artificial}.

Despite that the majority of the above-discussed DL methods focus on the cine CMR reconstruction, they should be generalizable to reconstruct each multi-contrast CMR in T1/T2 mapping. Alternatively, T1/T2 values can also be reconstructed directly without recovering each contrast image, such as using a fully connected neural network to predict the values directly~\citep{guo2022accelerated}. However, the current literature on CMR T1/T2 mapping reconstruction is limited, which could be explained by the lack of public CMR T1/T2 mapping datasets. Our effort in putting forward this CMRxRecon challenge with both cine and T1/T2 mapping CMR data will likely further strengthen the active research in the field.

\section{Challenge Setup}

\subsection{Dataset}

\subsubsection{Dataset Information}
Our institutional review board granted approval for the study (approval number: FE20017). Data collection was conducted using a 32-channel specialized cardiac coil in conjunction with a 3T scanner (MAGNETOM Vida, Siemens Healthineers, Germany). We successfully recruited and acquired data from 300 healthy volunteers at our medical center. The data are divided into three sets, i.e., 120 training data, 60 validation data and 120 test data.
Before the scans, participants were positioned in a supine posture. 
During the scan, electrodes were connected, and an electrocardiogram (ECG) signal was recorded. 
Cardiac Scout imaging was conducted using the `Dot' engine. 
The CMRxRecon dataset~\citep{wang2023cmrxrecon} follows the CC-BY license. All released data provided by the challenge is publicly available but limited to non-commercial use. 

We followed the CMR imaging protocol given in the earlier publication~\citep{wang2021recommendation}. 
For 2D cardiac cine, the "TrueFISP" readout was employed. Long-axis (LAX) and short-axis (SAX), two-chamber (2CH), three-chamber (3CH), and four-chamber (4CH) views were gathered. 
Generally, 5 $\sim$ 11 slices were obtained for the SAX view while a single slice was obtained for each of the other views. Using a temporal resolution of around 50 ms, the cardiac cycle was divided into 12~25 phases based on heart rate. 
Typical scan settings included an 8.0 mm slice thickness, a 1.5 $\times$ 1.5 mm$^2$ spatial resolution, a 3.6 ms repetition time (TR), and a 1.6 ms echo time (TE). 
The original acceleration factor for parallel imaging was R = 3. The signal was obtained with breath-holding. 

Using a modified look-locker inversion recovery (MOLLI) sequence, nine images with various T1 weightings (using the 4-(1)-3-(1)-2 scheme) were obtained for T1 mapping. 
T1 mapping was performed only in SAX view, with a typical field-of-view (FOV) of 340 $\times$ 340 mm$^2$, spatial resolution of 1.5 $\times$ 1.5 mm$^2$, slice number of 5 or 6, slice thickness of 5.0 mm, TR of 2.7 ms, TE of 1.1 ms, partial Fourier of 6/8, and parallel imaging acceleration factor of R = 2. 
Subjects' inversion times differed based on their heart rates in real-time. Using an ECG trigger, signals were obtained after the diastole. 

Using T2-prepared (T2prep)-FLASH sequence and three T2 weightings in SAX view, T2 mapping was carried out using the same geometrical parameters as T1 mapping and similar imaging parameters, including 
340 $\times$ 340 mm$^2$ FOV, 1.5 $\times$ 1.5 mm$^2$ spatial resolution, 5$\sim$6 slices, 5.0 mm slice thickness, 3.0 ms TR, 1.3 ms TE, 0/35/55 ms T2 preparation time, 6/8 partial Fourier, and R = 2 parallel imaging acceleration factor. 

All the images were reconstructed using GeneRalized Autocalibrating Partially Parallel Acquisitions (GRAPPA). For the purpose of challenge, we retrospectively undersampled the k-space data with acceleration factors of R = 4, 8, 10 with uniform sampling trajectory.

\subsubsection{Annotation Details}
The myocardium and chambers were manually segmented by a skilled radiologist with over 5 years of expertise in cardiac imaging using ITK-SNAP (version 3.8.0).
The original image coordinates were preserved in NIFTI format together with the segmentation labels and matching images. 
The following four chamber labels apply to the LAX cine images: 
\begin{itemize}
    \item[] a) Label 1 for the left atrium; 

    \item[] b) Label 2  for the right atrium;

    \item[] c) Label 3 for the left ventricle; 

    \item[] d) Label 4 on the right ventricle. 
\end{itemize}
We labeled the SAX cine images using the following definitions:
\begin{itemize}
    \item[] a) Label 1 for the left ventricle blood pool;

    \item[] b) Label 2 for left ventricular myocardium;

    \item[] c) Label 3 for the right ventricle blood pool.
\end{itemize}
Both the T1 and T2 mapping annotations were identical to SAX cine.

\begin{figure*}[h]
\centering
\includegraphics[scale=.25]{ 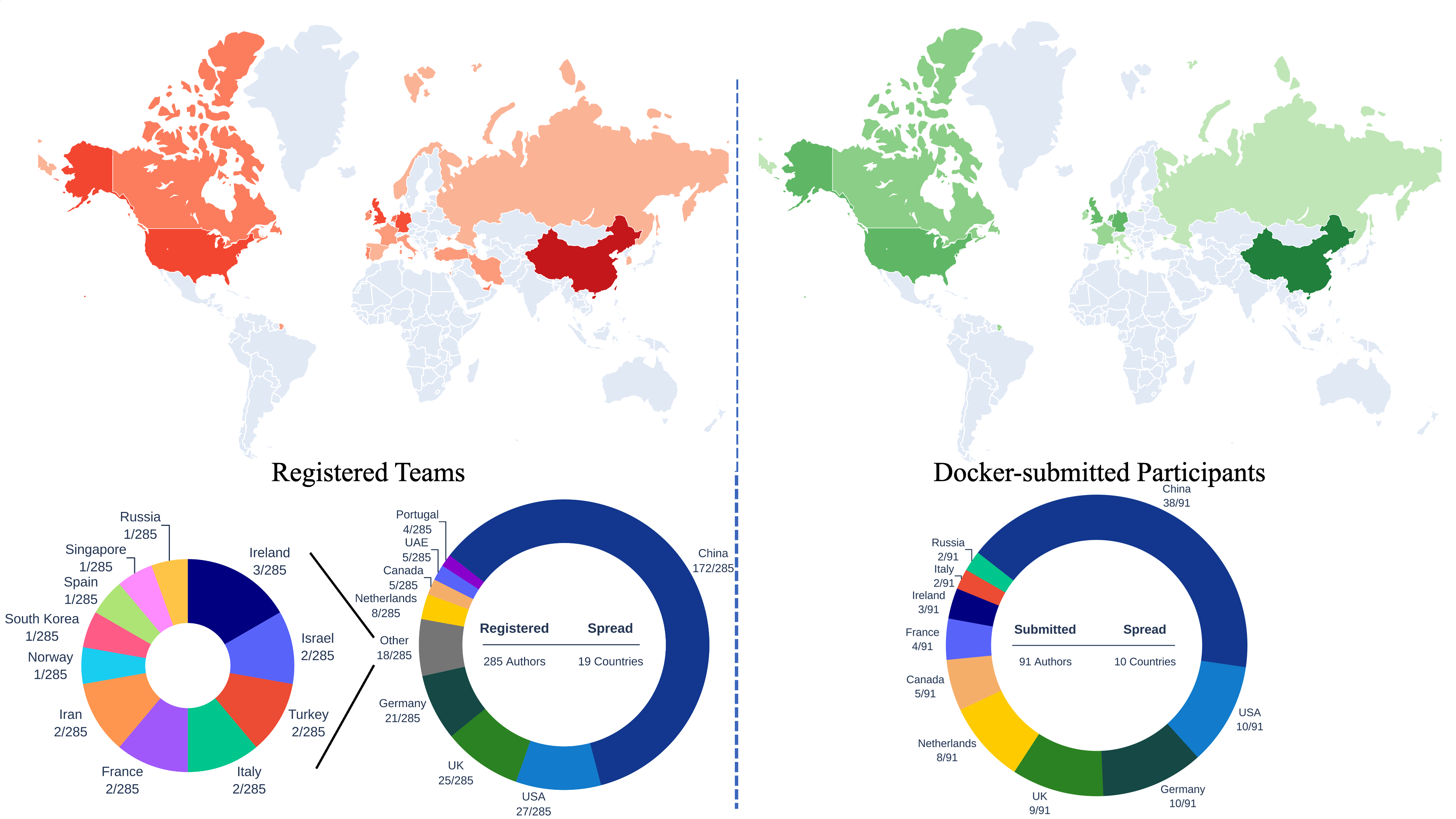}
\caption{The individual team statistics that registered and submitted the dockers for testing in the CMRxRecon Challenge.}
\label{fig:participant}
\end{figure*}

\begin{table*}[t]
\centering
\caption{The list and details of the participants and teams who successfully participated in the test (docker-submission) phase.}
\begin{tabular}{|p{3cm}|p{9.5cm}|p{3.5cm}|} \toprule
\small
 \textbf{Team name}     & \textbf{Affiliation}       & \textbf{Location}        \\ \hline \midrule
C1/M1. hellopipu     & Department of Computer Science, Rutgers University                                                              & New Brunswick, USA             \\ \hline
C2/M2. DIRECT        & AI for Oncology, Netherlands Cancer Institute                                                                & Amsterdam, Netherlands \\ \hline
C3/M3. clair         & Department of Imaging and Interventional Radiology, Faculty of Medicine, The Chinese University of Hong Kong & Hong Kong, China       \\ \hline
C4. tjubiit        & Tianjin Key Laboratory of Brain Inspired Intelligence Technology (BIIT), Tianjin University                  & Tianjin, China           \\ \hline
M4. dbmapping     & Department of Imaging Physics, Delft University of Technology                     &Delft, Netherlands \\ \hline
C5. imr            & Canon Medical Systems (China) Co., Ltd.                    & Beijing, China           \\  \hline
C6/M5. jabbers       & Physikalisch-Technische Bundesanstalt (PTB)                               & Berlin, Germany         \\ \hline
M6. whitealbum2   & Nanjing University of Aeronautics and Astronautics                                                           & Nanjing, China           \\ \hline
C7/M7. SkoICIG       & Moscow, Skolkovo Institute of Science and Technology                                                        & Russia          \\ \hline
C8. mataffine               & School of Artificial Intelligence, Beijing Normal University                                                 & Beijing, China\\ \hline
M8/C11. Fast2501      & Department of Biomedical Engineering, University of Virginia                                                 & Charlottesville, USA             \\ \hline
C9. OREO & Beijing University of Posts and Telecommunications                                                           & Beijing, China \\ \hline
M9. imperial\_cmr & National Heart and Lung Institute, Imperial College London                                               & London, UK  \\ \hline
C10. flyer                   & Department of Radiation Oncology, Peking University Cancer Hospital \& Institute                             & Beijing, China  \\ \hline
M10/C17. IADI-IMI      & Institute of Medical Informatics, University of Lübeck; IADI, Inserm U1254                                       & Lübeck, Germany; Nancy, France         \\ \hline
C12. Edipo                   & School of Engineering,  University of Edinburgh; Department of Information Engineering, University of Pisa                                       & Edinburgh, UK; Pisa, Italy \\\hline
C13. hkforest                & Electronic and Computer Engineering, the Hong Kong University of Science and Technology                      & Hong Kong, China  \\ \hline
M11. sunnybrook    & Department of Medical Biophysics, University of Toronto                                                      & Toronto, Canada          \\  \hline
C14. tsinghuacbir            & Tsinghua University                                                                                          & Beijing, China           \\ \hline
C15. insightdcu              & Insight SFI Research Centre for Data Analytics, Dublin City University                                       &Dublin, Ireland         \\ \hline
C16. lyulab                  & Shenzhen Technology University                                                                               & Shenzhen, China           \\ \hline
C18. fzu312lab               & Biomedical Engineering Institute of Fuzhou University                                                        & Fuzhou, China           \\ \bottomrule
\end{tabular}
\label{tab:particpant}
\end{table*}

\subsection{Participants}

The CMRxRecon Challenge is held in conjunction with the 26th International Conference on Medical Image Computing and Computer-Assisted Intervention (MICCAI 2023), on October 12th, 2023 in Vancouver, Canada. The official website for the CMRxRecon challenge is (\href{https://cmrxrecon.github.io/}{https://cmrxrecon.github.io}).As an open challenge, CMRxRecon received a total of 285 registration requests before the deadline preceding the MICCAI 2023 conference. 
Among them, 22 teams with 91 participants successfully submitted algorithm Docker containers for the testing phase before the submission deadline, as illustrated in Figure~\ref{fig:participant}. Among them, 7 teams submitted for both cine and mapping tasks.
The 91 participants represent a diverse cohort hailing from 10 different countries. 
The details of all participating teams are summarized in Table~\ref{tab:particpant}. 
Note that a unique team index was assigned to each team participating in different tasks. 
For simplicity, we denote teams engaged in cine reconstruction as `C$X$' and those involved in mapping reconstruction as `M$X$'. 
We have carefully selected and extensively reported 10 representative algorithms, which included the results from the top 5 teams as well as the five teams that our organizers unanimously considered to be the most distinctive. The chosen algorithms take into account both novelty and performance evaluation.
All teams have given their consent for the inclusion of their methods and results in this publication. 

\subsection{Challenge Phases}
The challenge includes three phases. 
First, to complete the registration and get access to the training and validation dataset, the participants were requested to register on the official challenge website, sign the data agreement, and keep their promise to abide by the challenge rules. 
Second, the participants were invited to take part in the validation phase, where the reconstructed images from under-sampled data are required to be submitted. The evaluation is automatically executed on the Synapse platform \url{https://www.synapse.org/#!Synapse:syn51471091/wiki}. The leaderboard is also presented online and updated promptly. 
Third, the participants were invited to take part in the final test phase to complete the full participation in this challenge. To guarantee the fairness of the competition, the packaged docker is the only valid submission in the test stage. Each team can submit 3 docker containers, and we will take the final submission as the official one. Our staff will run the docker container and confirm with the team that it has been successfully executed. Only completing the predictions of all test cases will be considered successful participation. The prizes are awarded to the top-3 teams of each task during the MICCAI conference. The top-3 teams in each task were invited to report their methodologies in the STACOM workshop on October 12, 2023.

\subsection{Evaluation Metrics}
The reconstruction performance for both cine and mapping were assessed using the following criteria: peak signal-to-noise ratio (PSNR), normalized mean square error (NMSE), and structural similarity index measure (SSIM). For T1 and T2 mapping, we also calculated the quantitative T1 and T2 relaxation times in  myocardium for comparison. The root mean square error (RMSE) for T1 and T2 values in myocardium was computed as evaluation metrics as well.
We use the SSIM as our quantitative  quality metric for ranking. For the cases without valid output, we will assign it to the lowest value of metric.

The metrics were defined as follows:

\noindent \textbf{SSIM} The SSIM index utilizes the inter-pixel relationships to assess the similarity between two images.
The resemblance that results between two image patches, $\hat{m}$ and $m$, is described as follows.
\begin{equation}
\operatorname{SSIM}(\hat{m}, m)=\frac{\left(2 \mu_m \mu_m+c_1\right)\left(2 \sigma_{m m}+c_2\right)}{\left(\mu_{\hat{m}}^2+\mu_m^2+c_1\right)\left(\sigma_{\hat{m}}^2+\sigma_m^2+c_2\right)^{\prime}}
\end{equation}
where $c_1$ and $c_2$ are two variables to stabilize the division; $c_1=\left(k_1 L\right)^2$ and $c_2=\left(k_2 L\right)^2$. $\mu_{\hat{m}}$ and $\mu_{{m}}$ are the average pixel intensities in $\hat{m}$ and $m$, and their variances are $\sigma_{\hat{m}}^2$ and $\sigma_{{m}}^2$.

\noindent \textbf{PSNR} The power of the highest image intensity that may be achieved across a volume divided by the power of distortion-causing noise and other defects is known as the PSNR.

\begin{equation}
\operatorname{PSNR}(\hat{v}, v)=10 \log _{10} \frac{\max (v)^2}{\operatorname{MSE}(\hat{v}, v)}.
\end{equation}
Let $v$ represent the target volume, $\hat{v}$ represent the reconstructed volume, $max(v)$ denote the largest entry in the target volume, and MSE$\hat({v},v)$ be the mean square error between $\hat{v}$ and $v$, which is defined as $\frac{1}{n}|\hat{v}-v|_2^2$. Here, $n$ represents the total number of entries in the target volume $v$.

\noindent \textbf{NMSE} The NMSE between a reference image or volume $v$ and a reconstructed image or image volume expressed as a vector $\hat{v}$ is defined as:
\begin{equation}
\operatorname{NMSE}(\hat{v}, v)=\frac{\|\hat{v}-v\|_2^2}{\|v\|_2^2},
\end{equation}
where the squared Euclidean norm is represented by $\|\cdot\|_2^2$, and the subtraction is carried out entry-wise. 
We reported the NMSE values for the whole image volumes.

The Python scripts utilized for evaluating the image quality metrics can be found on GitHub at the following URL: \href{https://github.com/CmrxRecon/CMRxRecon/tree/main/Evaluation}{https://github.com/CmrxRecon/CMRxRecon/tree/main/Evaluation}.

\section{Comparative Overview on Participants Methodologies}
This section provides a comprehensive comparison across various methodologies. A detailed summary of the 10 selected approaches is presented in Table~\ref{table:team-strategies}. Within this table, we focus on delineating the principal contributions and the training strategies employed by the teams, providing a clear insight into the diverse techniques and their unique strengths. All participants trained their models from scratch on the CMRxRecon dataset.

\begin{table*}
\centering
\caption{Summary of the strategies and contributions of the 10 selected teams}

\scalebox{0.62}{
\begin{tabular}{|p{1.2cm}|p{8cm}|p{17cm}|}
\hline
\textbf{Team name} & \textbf{Main novelty/contribution} & \textbf{Training strategy} \\\\[-3ex]
\toprule
C1/M1. hellopipu & 
\vspace{-2mm}
\begin{itemize}[leftmargin=*, topsep=0pt, partopsep=0pt, itemsep=0pt, parsep=0pt]
    \item Use E2E-VarNet\cite{sriram2020end} as backbone, introduce PromptMR, a prompting-based all-in-one unrolled model
    \item Incorporate an image domain denoiser, PromptUnet, coupled with a k-space domain data consistency layer 
\end{itemize}  &
\vspace{-2mm}
\begin{itemize}[leftmargin=*, topsep=0pt, partopsep=0pt, itemsep=0pt, parsep=0pt]
  \item Data: multi-coil cine and mapping
  \item Input: complex image data separated into 2-channel of real and imaginary parts
  \item Normalization: z-score normalization
  \item Learning: AdamW optimizer ($\beta1$=0.9, $\beta2$=0.999, weight decay=0.01) over 12 epochs, with an initial learning rate of \( 1 \times 10^{-4} \) and decayed to \(1 \times 10^{-5}\) in the last epoch
  \item Loss: SSIM 
  \item Code: \url{https://github.com/hellopipu/PromptMR}
\end{itemize} \\[-2ex]
\hline
C2/M2. DIRECT & 
\vspace{-2mm}
\begin{itemize}[leftmargin=*, topsep=0pt, partopsep=0pt, itemsep=0pt, parsep=0pt]
    \item Use vSHARP \cite{yiasemis2023vsharp} as backbone, customize it to a 3D variant tailored specifically for 2D dynamic reconstruction;
    \item Comprise three key steps: a denoising step to refine the auxiliary variable, a data consistency step for the target image performed through differentiable gradient descent over Tx=8 iterations, and an update for the Lagrange Multipliers introduced by ADMM 
\end{itemize}  & 
\vspace{-2mm}
\begin{itemize}[leftmargin=*, topsep=0pt, partopsep=0pt, itemsep=0pt, parsep=0pt]
    \item Data: multi-coil cine and mapping
    \item Input: complex data was separated into a 2-channel format
    \item Normalization: scaling the initial k-space with the 99.5\% percentile value of its modulus
    \item Augmentation: joint modality training (cine and mapping data concurrently) and augmentations like k-space flipping, random k-space cropping, and multi-scheme undersampling introduced \cite{yiasemis2022direct}
    \item Learning: Adam optimizer ($\beta1$ = 0.9, $\beta2$ = 0.999, $\epsilon = 10 ^{-8}$), initial linear increase to 0.003 over 2k iterations, followed by a reduction every 50k iterations by a factor of 0.9
    \item Loss: SSIM, SSIM3D, HFEN, and L1 losses in the image domain along with NMAE and NSME losses in the k-space domain
    \item Code: \url{https://github.com/NKI-AI/direct}
\end{itemize} \\[-2ex]
\hline
C3/M3. clair & 
\vspace{-2mm}
\begin{itemize}[leftmargin=*, topsep=0pt, partopsep=0pt, itemsep=0pt, parsep=0pt]
    \item Use CAMP-Net \cite{zhang2023camp} as backbone, propose k-t CLAIR \cite{zhang2023k} incorporates self-consistency guidance and multiple priors in deep learning to exploit spatio temporal correlations across the x-t, x-f, and k-t domains
\end{itemize} &
\vspace{-2mm}
\begin{itemize}[leftmargin=*, topsep=0pt, partopsep=0pt, itemsep=0pt, parsep=0pt]
    \item Data: multi-coil cine and mapping
    \item Dataset: 96 and 24 cases for training and validation 
    \item Learning: Adam optimizer ($\beta1 = 0.9$ and $\beta2 = 0.999$) 30/50 epochs with learning rate: $3 \times 10^{-4}$ at initial and were reduced by a factor of 10 in the last 10 epochs
    \item Loss: SSIM and L1 
    \item Code: \url{https://github.com/lpzhang/ktCLAIR}
\end{itemize} \\[-2ex]
\hline
M4. dbmapping &
\vspace{-2mm}
\begin{itemize}[leftmargin=*, topsep=0pt, partopsep=0pt, itemsep=0pt, parsep=0pt]
    \item Use unrolling gradient descent scheme~\citep{hammernik2018learning} as the backbone for multi-coil network
    \item Introduce a relaxometry-informed quantitative MRI reconstruction method that synergizes joint mapping and unrolled gradient descent reconstruction,
\end{itemize} &
\vspace{-2mm}
\begin{itemize}[leftmargin=*, topsep=0pt, partopsep=0pt, itemsep=0pt, parsep=0pt]
    \item Data: multi-coil mapping
    \item Normalization: The k-space data were segmented into their real and imaginary components, then normalized by the value at the 99th percentile of the frequency domain magnitude
    \item Augmentation: random Gaussian noise
    \item Data consistency: coil sensitivity maps were initially estimated by SENSE~\citep{pruessmann1999sense} for multi-coil data
    \item Learning: Adam optimizer for 400 epochs with an initial learning rate $10^{-3}$ and polynomial weight decay
    \item Loss: L1, SSIM and fitting loss in mapping
    \item Code: \url{https://github.com/pandafriedlich/relax_qmri_recon}
\end{itemize}\\[-2ex]
\hline
C4. tjubiit & 
\vspace{-2mm}
\begin{itemize}[leftmargin=*, topsep=0pt, partopsep=0pt, itemsep=0pt, parsep=0pt]
    \item Use soft data-consistency \cite{sriram2020end} as the backbone and showcase a novel multi-scale inter-frame information fusion strategy
    \item Integrate distinct encoders dedicated to extracting features from each frame. These extracted features were then combined effectively using an information fusion block that incorporated multi-scale features from multiple frames
\end{itemize} & 
\vspace{-2mm}
\begin{itemize}[leftmargin=*, topsep=0pt, partopsep=0pt, itemsep=0pt, parsep=0pt]
    \item Data: multi-coil cine
    \item Dataset: 120, 60, and 120 cases for training, validation, and test
    \item Input: the real and imaginary components of the data were concatenated along the channel dimension
    \item Augmentation: spatial domain data underwent random vertical and horizontal flipping and rotation at an angle not exceeding 45°, each with a defined probability
    \item Learning: the initial learning rate for training was set to 0.005 and was scheduled to decay by a factor of 0.1 every 40 epochs 
    \item Loss: SSIM
\end{itemize} \\[-2ex]
\hline
C7/M7. Skolcig & 
\vspace{-2mm}
\begin{itemize}[leftmargin=*, topsep=0pt, partopsep=0pt, itemsep=0pt, parsep=0pt]
    \item Parameterize the objective function in compressed sensing (CS) minimization procedure by deep learning model \cite{kuzmina2022autofocusing+} as backbone
    \item Use meta-learning for CS minimization; for multi-coil, GRAPPA \cite{griswold2002generalized} is first used initial estimation and used a simple U-net \cite{falk2019u} to solve CS problem.
\end{itemize}  & 
\vspace{-2mm}
\begin{itemize}[leftmargin=*, topsep=0pt, partopsep=0pt, itemsep=0pt, parsep=0pt]
    \item Data: multi-coil and single-coil cine and mapping
    \item Learning: Adam optimizer ($\beta1$ = 0.9 and $\beta1$ = 0.999) and learning rate $10^{-3}$
    \item Loss: L1
    \item Code: \url{https://github.com/Airplaneless/cmrx}
\end{itemize} \\[-2ex]
\hline
M8. Fast2501 & 
\vspace{-2mm}
\begin{itemize}[leftmargin=*, topsep=0pt, partopsep=0pt, itemsep=0pt, parsep=0pt]
    \item Use a 2D U-Net \cite{falk2019u} as the backbone network structure and propose a complex-valued cascading cross-domain convolutional neural network
    \item Alternate between the restoration step and the data consistency step
\end{itemize} &
\vspace{-2mm}
\begin{itemize}[leftmargin=*, topsep=0pt, partopsep=0pt, itemsep=0pt, parsep=0pt]
    \item Data: multi-coil cine and mapping
    \item Dataset: 90,10,20 cases for training, validation, and testing.
    \item Normalization: k-space data for each 2D slice was scaled to have its magnitude between 0 to 1
    \item Augmentation: random flipping along readout and phase encoding directions was employed as training augmentation. During training, the undersampling ratio was randomly selected between 4 to 12, and the equispaced undersampling mask was generated on the fly.
    \item Learning: Adam optimizer for 50 epochs with a learning rate of 0.0001, $\beta1$ = 0.9, $\beta2$ = 0.999, and $\epsilon = 10^{-8} $
    \item Loss: L1 and SSIM
\end{itemize} \\[-2ex]
\hline
C12. Edipo & 
\vspace{-2mm}
\begin{itemize}[leftmargin=*, topsep=0pt, partopsep=0pt, itemsep=0pt, parsep=0pt]
    \item Use convolutional recurrent neural network(CRNN) \cite{qin2018convolutional} and single-image super-resolution network, Bicubic++ \cite{bilecen2023bicubic++}, as the backbone
    \item Propose an additional bidirectional convolutional recurrent unit (BCRNN) followed by a lightweight refinement module.
\end{itemize}  & 
\vspace{-2mm}
\begin{itemize}[leftmargin=*, topsep=0pt, partopsep=0pt, itemsep=0pt, parsep=0pt]
    \item Data: single-coil cine
    \item Dataset: 90, 20, and 10 cases for training, validation, and testing
    \item Input: complex data are separated into magnitude and phase channels
    \item Strategy: jointly trained the short-axis (SA) and long-axis (LA) data during the training phase
    \item Learning: Adam optimizer($\beta1$=0.9, $\beta1$=0.999, weight decay 0.0) for 50 epochs with the initial learning rate $3 \times 10^{-4}$ gradually reduced to $3 \times 10^{-6}$ with the StepLR scheduler.
    \item Loss: L1 and SSIM
    \item Code: \url{https://github.com/vios-s/CMRxRECON_Challenge_EDIPO}
\end{itemize} \\[-2ex]
\hline
C15. insightdcu & 
\vspace{-2mm}
\begin{itemize}[leftmargin=*, topsep=0pt, partopsep=0pt, itemsep=0pt, parsep=0pt]
    \item Use Unet \cite{falk2019u} as the backbone, enhanced by Group Normalization (GN) and channel attention layers (Gated Channel Transformation: GCT) \cite{yang2020gated}
\end{itemize}  & 
\vspace{-2mm}
\begin{itemize}[leftmargin=*, topsep=0pt, partopsep=0pt, itemsep=0pt, parsep=0pt]
    \item Data: single-coil cine
    \item Dataset: random sampling strategy to obtain 1000 LAX and 1000 SAX images for training
    \item Learning: AdamW was used as an optimizer with a learning rate of 0.001
    \item Loss: MSE
    \item Normalization: scaled by the maximum signal intensity and further min-max scaled into the range of [0,1]
    \item Code: \url{https://github.com/juliadietlmeier/CMRxRecon_insightdcu}
\end{itemize} \\[-2ex]
\hline
C17. IADI-IMI & 
\vspace{-2mm}
\begin{itemize}[leftmargin=*, topsep=0pt, partopsep=0pt, itemsep=0pt, parsep=0pt]
    \item Use multi-resolution hash encoding \cite{muller2022instant} and JSENSE \cite{ying2007joint} for implicit sensitivity map estimation as backbone
    \item Consist of two shallow MLP networks for the simultaneous prediction of a complex-valued intensity reconstruction and a set of complex-valued coil sensitivity maps
\end{itemize}  & 
\vspace{-2mm}
\begin{itemize}[leftmargin=*, topsep=0pt, partopsep=0pt, itemsep=0pt, parsep=0pt]
    \item Data: multi-coil cine
    \item Dataset: 0 for training (instance-optimization) and 20 for testing
    \item Normalization: each under-sampled multi-coil 2D+t k-space was scaled by the maximum intensity of its SOS magnitude reconstruction.
    \item Learning: Adam optimizer ($\beta_1 = 0.9$, $\beta_2 = 0.99$) with a learning rate of $10^{-2}$ for 200 iterations
    \item Loss: Huber loss ($\delta =1.0$) and total-variation regularization
    \item Code: \url{https://github.com/MDL-UzL/CineJENSE}
\end{itemize}   \\[-1ex] \bottomrule \hline
\end{tabular}
}
\label{table:team-strategies}
\end{table*}

Next, we report the methods from those selected teams in detail and highlight the key novelty or contribution of each method.

\subsection{C1/M1 hellopipu}
The team of \emph{hellopipu} (C1/M1) proposed a two-stage MRI reconstruction pipeline to address the limitations of existing MRI reconstruction methods. 
\begin{figure}[!t]
\centering
\includegraphics[scale=.5]{ 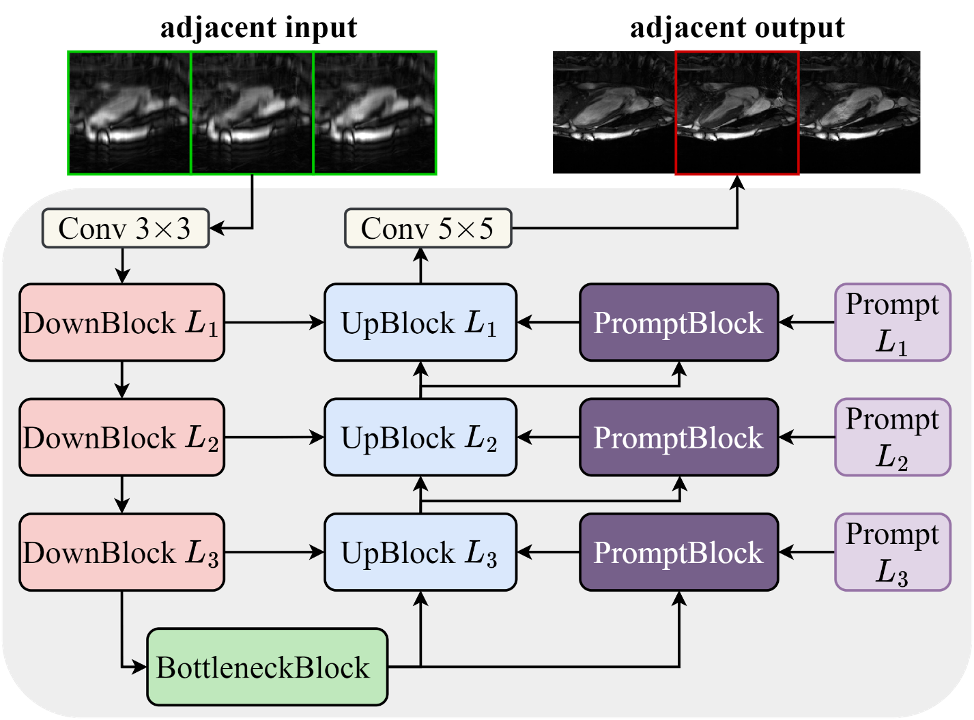}
\caption{ Overview of PromptUnet proposed by the team \emph{hellopipu} (C1/M1). PromptUnet serves as the denoiser in each cascade of PromptMR. It processes adjacent input to explore the inter-frame/-contrast information and incorporates a PromptBlock at each level to allow rich hierarchical context learning. }
\label{fig:hellopipu}
\end{figure}
Expanding on the foundation of E2E-VarNet~\citep{sriram2020end}, the researchers introduced PromptMR, a comprehensive unrolled model for MRI reconstruction based on prompting. This model is versatile for handling diverse views, contrasts, adjacent types, and acceleration factors present in real clinical cardiac MRI scans. Within the architecture of PromptMR, each cascade integrates an image domain denoiser called PromptUnet, as shown in Figure~\ref{fig:hellopipu}, and a k-space domain data consistency layer. PromptUnet employs a 3-level encoder-decoder architecture, featuring DownBlock, UpBlock, and PromptBlock at each level. The utilization of adjacent input~\citep{fabian2022humus} and channel attention in PromptUnet facilitates the exploration of inter-frame/-contrast information. The PromptBlock, inspired by PromptIR~\citep{potlapalli2023promptir}, encodes specific input-type context as an adaptively learned prompt across multiple levels to guide the reconstruction process. The multi-coil sensitivity maps are estimated by a compact PromptUnet from the central k-space which serves as the auto-calibration signal (ACS).


For training, both multi-coil SAX/LAX cine data and T1/T2-weighted data from the 120 healthy subjects in the dataset were employed. The input image to each PromptUnet in PromptMR was normalized using z-score normalization. Data augmentation during training focused on balancing the portion of SAX/LAX/T1/T2 slices. The complex image data was separated into 2 channels of real and imaginary parts as input to the network. A single PromptMR model with 12 cascades was trained. The model was optimized using the AdamW optimizer, configured with specific parameters: $\beta_{1}$ was set to 0.9, $\beta_{2}$ to 0.999, with a weight decay of 0.01. The training utilized SSIM loss across 12 epochs, starting with an initial learning rate of $10^{-4}$, which was then reduced to $10^{-5}$ during the last epoch. Training took approximately 3 days on two NVIDIA A100 40GB GPUs with a batch size of one per GPU. To refine the reconstruction results of PromptMR, two ShiftNet models~\citep{li2023simple} and test time augmentation by flipping and 180-degree rotation were incorporated in the second stage. Each ShiftNet was trained with the initial cine or mapping reconstructions by PromptMR as the input. We used the Adam optimizer ($\beta_{1}$=0.9, $\beta_{2}$=0.999, and weight decay=0), a batch size of one, cosine annealing learning rate schedule (base lr=\( 4 \times 10^{-4} \), $\eta_{min}$=\( 1 \times 10^{-7} \)), and SSIM loss to train ShiftNets for 50 epochs. The second stage of training took approximately 1 day for cine and 8 hours for mapping on 8 A100 40GB GPUs.
The codebase for this approach is available at \url{https://github.com/hellopipu/PromptMR}.

In summary, the design of PromptUnet plays a crucial role in leveraging adjacent k-space information and facilitating discriminative context learning for various MRI reconstruction tasks.

\subsection{C2/M2 DIRECT}
\begin{figure}[]
\centering
\includegraphics[scale=.07]{ 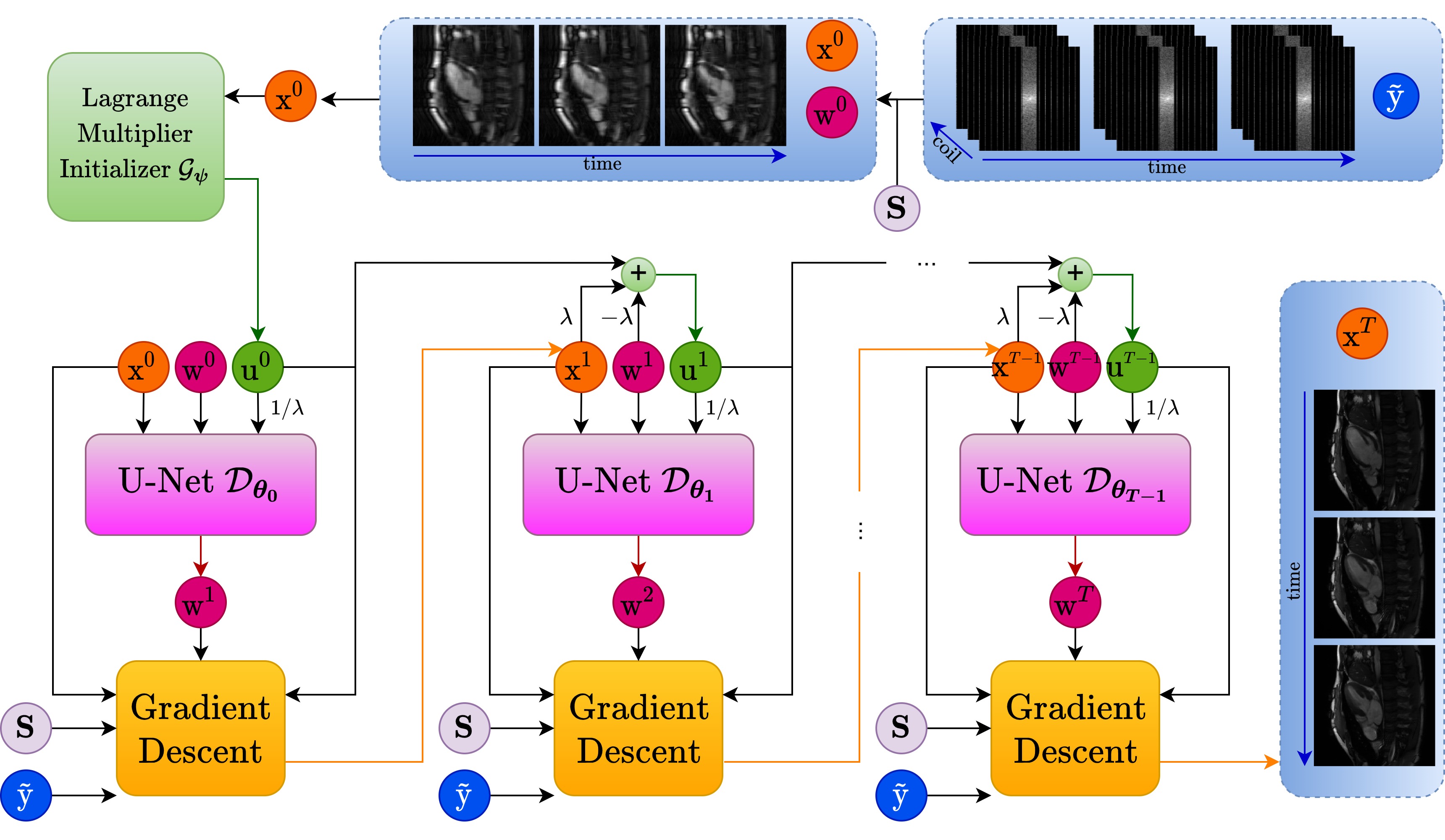}
\caption{Overall workflow of the proposed vSHARP by team \emph{DIRECT}(C2/M2).}
\label{fig:direct}
\end{figure}
The team of \emph{DIRECT} (C2/M2) formulated the reconstruction task as a least squares regularized optimization, with the adoption of vSHARP~\citep{yiasemis2023vsharp}, a variable Splitting Half-quadratic Alternating Direction Method of Multipliers (ADMM) algorithm~\citep{fukushima1992application} for the Reconstruction of Inverse-Problems, as the backbone.
The team optimized the method for both 2D and 3D (2D + time/contrast) reconstructions, although their submitted model comprised the 3D vSHARP variant.
This customized approach consists of three integral steps: first, a denoising step is implemented to refine the auxiliary variable; second, a data consistency step for the target image is executed through differentiable gradient descent over 8 iterations; and finally, an update mechanism for the Lagrange Multipliers is introduced by the ADMM. A distinctive feature of their approach is the integration of a Lagrange Multiplier Initializer, which utilizes a dilated convolution and replication padding module. This module, inspired by prior work \cite{recurrentvarnet} and adapted for 3D applications, generates an initial estimate for the Lagrange Multipliers.
The 2D dynamic vSHARP model took a sequence of 2D undersampled multi-coil k-space data (time-frames for cine tasks, contrast-frames for mapping tasks) as input and generated a corresponding sequence of 2D images as output for reconstruction. For each denoising step, distinct 3D U-Nets~\citep{ronneberger2015u} with four scales and 32 filters in the initial scale were employed. Due to the multi-coil nature of the data, a 2D U-Net (four scales, 32 channels in the first scale) was integrated for sensitivity map refinement from the ACS-k-space comprising 24 center lines.

The original complex data was separated into a 2-channel format as input to the network. To ensure robust training, normalization involved scaling the initial k-space with the $99.5\%$ percentile value of its modulus. Model performance was enhanced through techniques such as joint modality training (cine and mapping data concurrently) and augmentations like k-space flipping, random k-space cropping, and multi-scheme undersampling (radial, spiral, variable density, random and equispaced rectilinear following methods presented in ~\citep{yiasemis2024retrospective}). Additionally, the model simultaneously underwent training for all acceleration factors (4, 8, and 10).
A dual-domain loss function was employed, encompassing SSIM, SSIM3D, HFEN, and L1 losses in the image domain, along with NMAE and NSME losses in the k-space domain. The end-to-end pipeline underwent training for approximately 1 million iterations (around 25 days) using the Adam optimizer ($\beta1$ = 0.9, $\beta2$ = 0.999, $\epsilon = 10^{-8}$) for parameter optimization. The learning rate underwent an initial linear increase to 0.003 over 2k iterations, followed by a reduction every 50k iterations by a factor of 0.9. Training was conducted on four NVIDIA A100 80GB GPUs, with a batch size of 1 on each GPU.
They trained their proposed method on all provided training data and evaluated it on the validation set using the tool provided by the challenge. The time required for the reconstruction of a single 4D (space dimensions + time/contrast) volume varied between 2.7 to 15.7 seconds.
The code for this training strategy is openly accessible at \url{https://doi.org/10.21105/joss.04278}.

\subsection{C3/M3 clair}
\begin{figure}[]
\centering
\includegraphics[scale=.5]{ 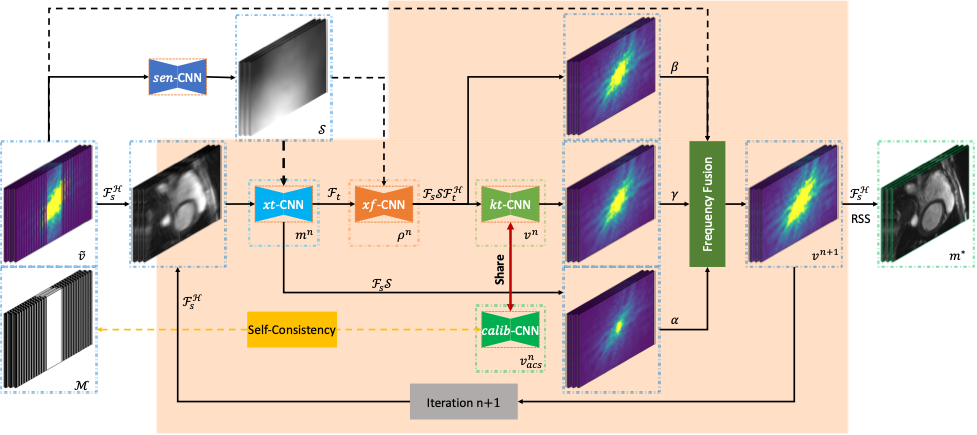}
\caption{The proposed k-t CLAIR by team \emph{clair}(C3/M3) exploits spatiotemporal correlations in data and incorporates calibration information to learn complementary priors across the x-t, x-f, and k-t domains.}
\label{fig:clair}
\end{figure}
The team \emph{clair}(C3/M3) introduced k-t CLAIR which adopted the unrolled-based CAMP-Net~\citep{zhang2023camp} as the foundational framework and expanded its capabilities to address dynamic and parametric CMR, as described in Figure~\ref{fig:clair}. 
By exploiting spatiotemporal correlations, k-t CLAIR learns complementary priors in the x-t, x-f, and k-t domains, while enforcing self-consistency learning in the k-t domain. 
The approach involves four key steps within each iteration: image enhancement in the x-t domain using xt-CNN, dynamic temporal prior learning in the x-f domain through xf-CNN, k-space restoration in the k-t domain using kt-CNN, and self-consistency learning in the k-t domain via calib-CNN. 
During each iteration, the approach outlines the reconstruction steps in the x-t, x-f, and k-t domains, leveraging spatiotemporal correlations and periodic cardiac motion for effective dynamic feature restoration. 
A frequency fusion block is introduced to coordinate feature learning processes, and joint learning of coil sensitivity maps with sen-CNN enhances the reconstruction process. 
Calibration information is integrated into the k-t domain to ensure accurate signal restoration. 
U-Net is utilized for highly nonlinear prior learning, and a frequency fusion layer balances contributions from different priors. The frequency fusion block facilitates the coordination of different priors, contributing to accurate and faithful dynamic MRI reconstruction.

Distinct models were trained for multi-coil cine and T1/T2 mapping data. 
The training utilized 80\% of the 120 healthy subjects, while the remaining 20\% were reserved for model validation. 
An additional 60 healthy subjects were included for online testing, and no data processing or augmentation, except for data standardization, was applied. The original complex data was separated into two channels: phase and magnitude, serving as inputs to the networks. Model optimization employed the Adam optimizer ($\beta1 = 0.9$ and $\beta2 = 0.999$) along with SSIM and L1 losses for 30/50 epochs for the Cine/Mapping task, initiating with a learning rate of $3 \times 10^{-4}$ and a batch size of 1. Learning rates were reduced by a factor of 10 in the last 10 epochs. The number of iterations was set to 12 for all rolling-based models. The GitHub repository is accessible at: \url{https://github.com/lpzhang/ktCLAIR}

\subsection{M4 dbmapping}
\begin{figure}[]
\centering
\includegraphics[scale=0.9]{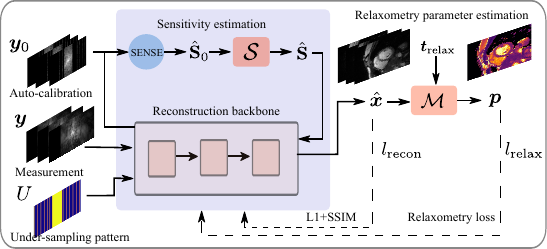}
\caption{The relaxometry-informed quantitative MRI reconstruction method that synergizes joint mapping and unrolled gradient descent reconstruction developed by the team \emph{dbmapping}(M4).}
\label{fig:dbmapping}
\end{figure}
The team \emph{dbmapping}(M4) proposes a relaxometry-guided reconstruction pipeline for the quantitative mapping subtask. Quantitative MRI reconstruction differs from other types of MRI in the sense that the reconstructed images should conform to the relaxometry. Taking advantage of this additional relaxometry prior, they proposed a joint mapping and reconstruction framework and employed an unsupervised mapping network to estimate the relaxometry-related parameters like T1 and T2. The design of the reconstruction backbone follows the unrolling gradient descent scheme~\citep{hammernik2018learning}. In each layer, the data fidelity is constrained in the frequency domain, and a U-Net is dedicated to learning the image prior in a data-driven fashion. They used exclusively the multi-coil acquisitions for reconstruction. Following~\citep{recurrentvarnet}, the coil sensitivity map was initially estimated by the SENSE~\citep{pruessmann1999sense} operator and then refined by a learnable U-Net. 

Multi-coil acquisitions were treated as image channels in the 2D reconstruction pipeline, and all baseline images were reconstructed simultaneously. The T1- and T2-mapping networks were trained to minimize the fitting loss induced by the relaxometry and estimate the quantitative parameters. Afterward, the mapping networks were frozen and incorporated into the reconstruction pipeline as relaxometry guidance. They trained separate neural networks for each acceleration factor (4/8/10) and each imaging sequence (MOLLI/T2-prep). The networks were trained in a 5-fold cross-validation manner using the Adam optimizer for 400 epochs, with an initial learning rate of $1 \times 10^{-3}$ and polynomial weight decay. A combination of L1, SSIM, and the fitting loss in mapping was used as the training loss. At inference time, an ensemble of the 5 models was used for the final prediction. 

In summary, they proposed a joint mapping and reconstruction framework for quantitative MRI and employed relaxometry and an additional prior for reconstruction. The GitHub repository is accessible at: \url{https://github.com/pandafriedlich/relax_qmri_recon}

\subsection{C4 tjubiit}
\begin{figure}[]
\centering
\includegraphics[scale=.9]{ 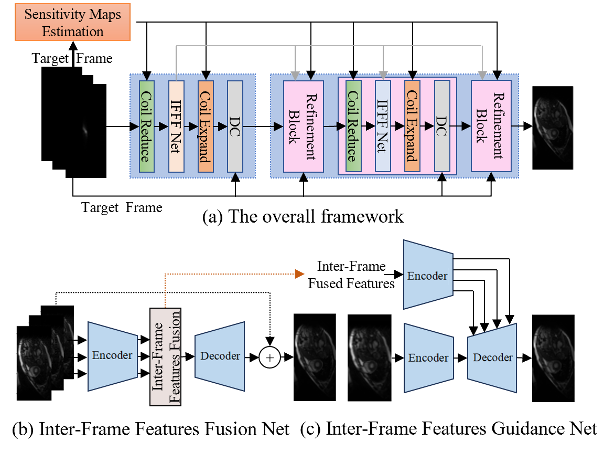}
\caption{The implementation of proposed Multi-Scale Inter-Frame Information Fusion Based Network by team \emph{tjubiit}(C4).}
\label{fig:tjubiit}
\end{figure}
The team \emph{tjubiit}(C4) presents an advanced approach to dynamic parallel MRI reconstruction, emphasizing a novel multi-scale inter-frame information fusion strategy, as shown in Figure~\ref{fig:tjubiit}. 
The method is designed to extract and leverage multi-scale features from adjacent multi-frame data, enhancing the overall reconstruction process. 
Specific encoders are employed for feature extraction from each frame, contributing to a nuanced understanding of information at different scales. 
The introduced information fusion block effectively combines these multi-scale features from multiple frames, enabling the comprehensive utilization of supplementary information. 
Additionally, the fused inter-frame information plays a crucial role in subsequent refinement blocks, guiding feature enhancement and contributing to the overall reconstruction quality. 
The proposed framework strategically incorporates several specific encoders for feature extraction from each frame in the inter-frame information fusion stage. 
This ensures a nuanced understanding of information at different scales, allowing for effective utilization of supplementary information from multiple frames. 
The information fusion block effectively combines the multi-scale features of multiple frames, facilitating comprehensive information utilization. 
Moreover, the fused inter-frame information is utilized in subsequent refinement blocks to enhance feature and guide the reconstruction process.
In the refinement stage, the method introduces an Inter-Frame Features Enhancement (IFFE) Net, which focuses on utilizing reference frame features for further enhancement. 
The IFFE Net adopts a U-Net architecture and introduces an Inter-Frame Features Enhancement Block (IFFEB) with spatial and channel attention mechanisms. 
Enhanced features are derived from a combination of spatial and channel attention maps, contributing to overall improvements in dynamic MRI reconstruction. 

The multi-coil cine k-space data, derived from the 120 subjects underwent division into 40\% for training, 20\% for validation, and 40\% for testing. 
The team employed two distinct data augmentation techniques. Specifically, spatial domain data underwent random vertical and horizontal flipping, as well as rotation at an angle not exceeding $45^{\circ}$, each with a defined probability. 
The real and imaginary components of the data were concatenated along the channel dimension and input into multiple cascaded networks. Different channels shared the same encoder-decoder network for estimating sensitivity maps from low-frequency data images. The framework ultimately integrated a soft data consistency layer~\citep{sriram2020end} to enhance fidelity. Following this, the frequency domain output underwent sequential processing involving inverse Fourier transform, absolute value calculation, and root sum squares calculation. Subsequently, supervised training was conducted using the SSIM loss. The initial learning rate for training was set to 0.005 and scheduled to decay by a factor of 0.1 every 40 epochs.

In conclusion, the introduced multi-scale inter-frame information fusion strategy not only significantly enhanced the overall reconstruction performance but also demonstrated a high level of efficiency.

\subsection{C7/M7 Skolcig}
\begin{figure}[]
\centering
\includegraphics[scale=.18]{ 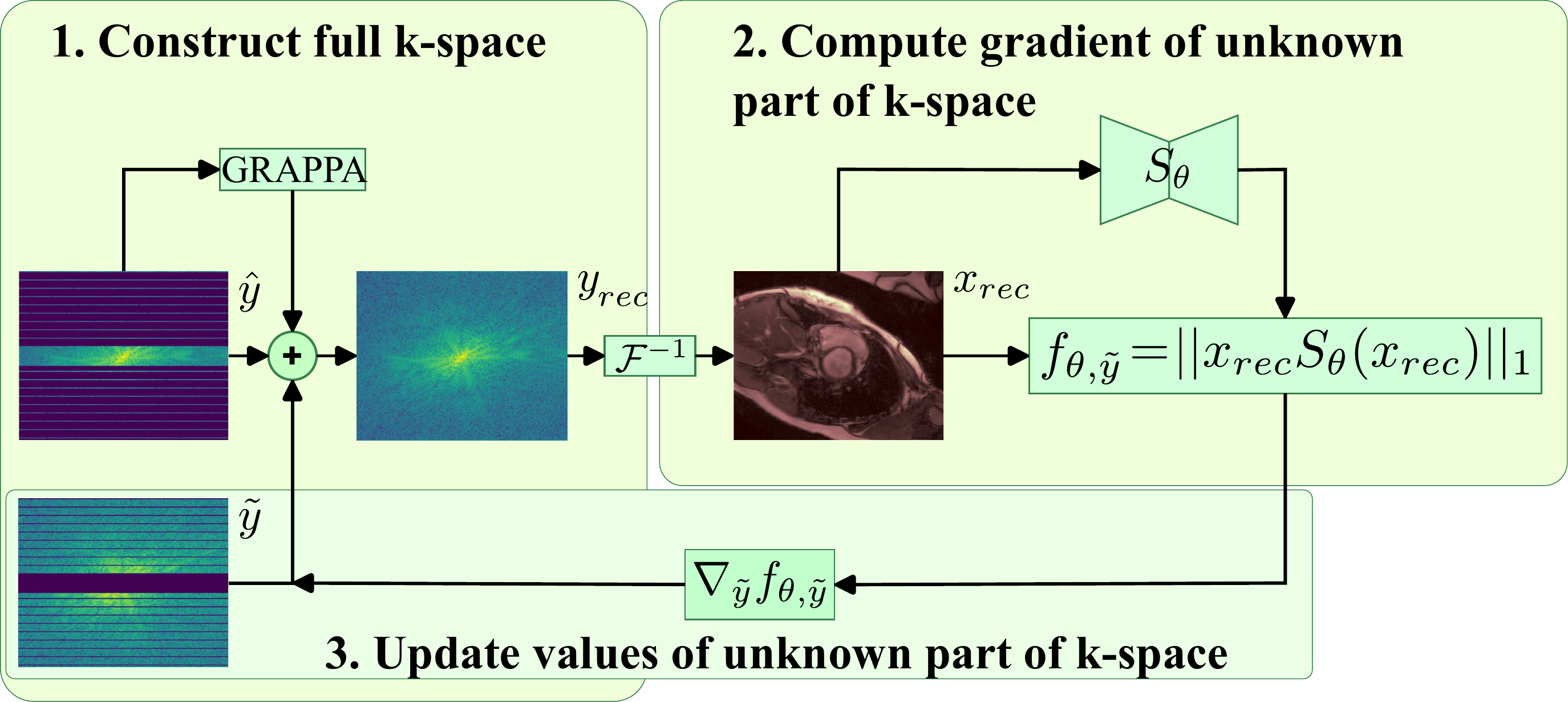}
\caption{Inference of the method of utilizing U-net like CNN for compressed sensing reconstruction developed by team \emph{Skolcig}(C7/M7). In the case of multi-coil data, $y_{rec} = \hat{y} + \tilde{y} + y_{grappa}$ where $y_{grappa}$ is the prediction of missing k-space data by GRAPPA convolution.}
\label{fig:Skolcig}
\end{figure}

The team \emph{Skolcig}(C7/M7) utilizes a novel approach by parameterizing the objective function in the compressed sensing (CS) minimization procedure through a deep learning model, specifically employing the model proposed in Autofocusing+~\citep{kuzmina2022autofocusing+} as a backbone. 
This can be regarded as meta-learning for CS minimization, as described in Figure~\ref{fig:Skolcig}. A commonly used objective function for compressed sensing is the L1-norm of the reconstructed image, denoted as $||x_{rec}||_1$. 
The SkolCIG team extended this function to $||x_{rec} S_{\theta, i}(x_{rec})||_1$, where $x_{rec}(\tilde{y}) = \text{rss}(\mathcal{F}^{-1}(\hat{y} + \tilde{y}))$ represents the estimation of the reconstructed image for a given sampled k-space $\hat{y}$, and $\tilde{y}$ is an estimate of unknown k-space data. $S_{\theta,i}(\cdot)$ denotes a U-net model with $\theta$ parameters on the $i$-th optimization step.

In the case of multi-coil k-space data, the estimation of the reconstructed image is given by $x_{rec}(\tilde{y}) = \text{rss}(\mathcal{F}^{-1}(\hat{y} + y_{grappa} + \tilde{y}))$, where $y_{grappa}$ is the estimation of the unknown part of the k-space using the GRAPPA method~\citep{griswold2002generalized}. The central 24 lines of k-space data were utilized for GRAPPA kernel estimation.
The CS minimization $\tilde{y} = \text{arg min} || x_{rec} S_{\theta,i}(x_{rec}) ||_1$ was performed using Adam optimization. The parameters of the U-net $\theta$ were also optimized by Adam optimization with parameters $\beta_1 = 0.9$, $\beta_2 = 0.999$, and a learning rate of $10^{-3}$. The SkolCIG team employed a simple U-net model~\citep{ronneberger2015u} with 32 channels and 4 pool layers, incorporating instance norm and SiLU activation.
Training such a U-net requires second-order gradients and storing the entire computational graph for CS optimization, making it a computationally and memory-intensive task. Therefore, they were constrained to 5 Adam optimization steps for $\tilde{y} = \text{arg min} || x_{rec} S_{\theta,i}(x_{rec}) ||_1$ minimization.
The implementation of this approach is available at \url{https://github.com/Airplaneless/cmrx}.

\subsection{M8 Fast2501}
\begin{figure}[]
\centering
\includegraphics[scale=.7]{ 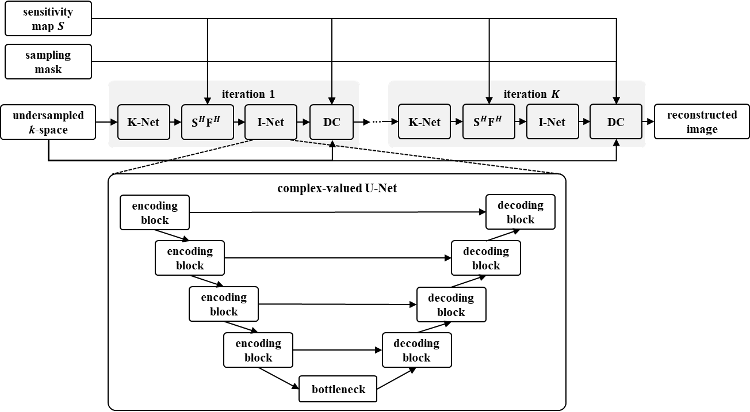}
\caption{The $C^3$-Net present by team \emph{Fast2501}(M8) alternates between the restoration step and the data consistency step. Both the k-space subnetwork and the image subnetwork use a complex-valued U-Net.}
\label{fig:fast2501}
\end{figure}
The team \emph{Fast2501}(M8) presents a sophisticated approach named $C^3$-Net—a complex-valued cascading cross-domain convolutional neural network designed for the reconstruction of undersampled CMR images, as shown in Figure~\ref{fig:fast2501}. The $C^3$-Net alternates between restoration and DC steps, incorporating a k-space subnetwork and an image subnetwork. Both subnetworks leverage a 2D U-Net~\citep{ronneberger2015u} as the backbone structure, featuring a sequence of complex-valued encoding or decoding blocks.

Separate models were trained for the cine and mapping tasks. The 120 fully sampled multi-coil subjects were randomly divided into 90 for training, 10 for validation, and 20 for testing. The magnitude of k-space data for each 2D slice was scaled to range from 0 to 1. Training augmentation included random flipping along readout and phase encoding directions. During training, the undersampling ratio was randomly chosen between 4 to 12, and the equispaced undersampling mask was generated dynamically. The central 24-phase encoding lines were consistently fully sampled, and sensitivity maps were pre-computed from the time-averaged autocalibration signal region using ESPIRiT~\citep{uecker2014espirit}. All subnetworks in the reconstruction pipeline underwent joint end-to-end training, utilizing a mixed L1 and SSIM loss. The training was conducted with an Adam optimizer for 50 epochs, employing a learning rate of 0.0001, $\beta1$ = 0.9, $\beta2$ = 0.999, $\epsilon = 10^{-8}$. The source code can be obtained from the corresponding author upon reasonable request.

In conclusion, the proposed C3-Net effectively integrates the complex value of MR data and coupled information from both the k-space domain and image domain within the model. This integration results in a substantial improvement in image quality, particularly at high acceleration rates.

\subsection{C12 Edipo}
\begin{figure}[]
\centering
\includegraphics[scale=.17]{ 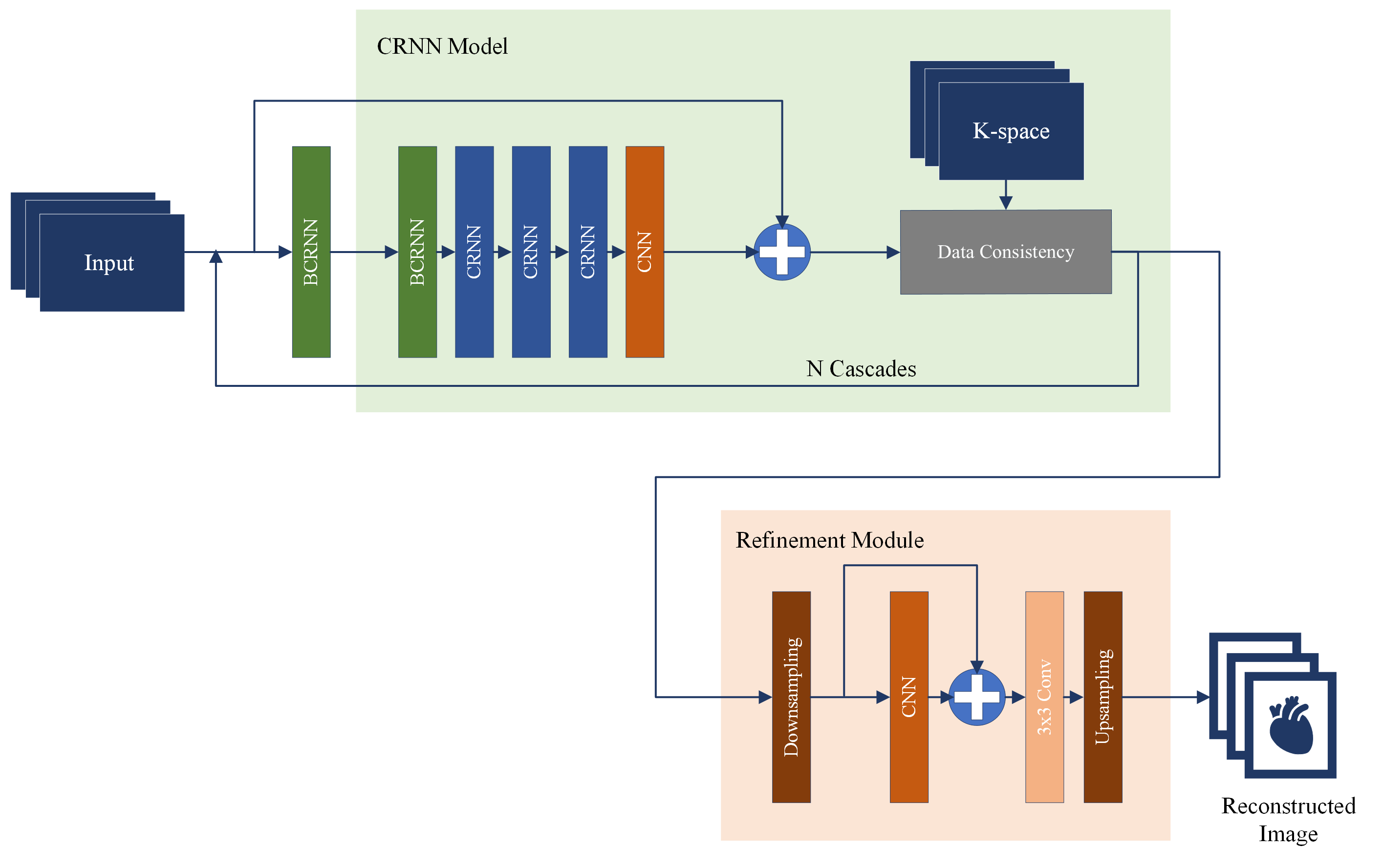}
\caption{The proposed model architecture by team \emph{Edipo}(C12): BCRNN, CRNN, and CNN units with a data consistency (DC) step for primary reconstruction. The low-cost refinement module includes downsampling (DS), CNN, and upsampling (US) units
}
\label{fig:edipo}
\end{figure}
The team \emph{Edipo}(C12) investigate the use of a convolutional recurrent neural network (CRNN)~\citep{qin2018convolutional} architecture and the single-image super-resolution network, Bicubic++~\citep{bilecen2023bicubic++} to exploit temporal correlations in supervised cine cardiac MRI reconstruction. In their proposed end-to-end network, as shown in Figure~\ref{fig:edipo}, they introduced an additional bidirectional convolutional recurrent unit (BCRNN) onto CRNN to specifically address motion artifacts by further exploiting spatio-temporal correlations. Following the CRNN module, a DC module was incorporated to enforce alignment between the reconstructed k-space and the lines acquired from the undersampled data. Lastly, a lightweight refinement module, inspired by super-resolution networks, was extended to enhance image details while maintaining a short reconstruction time. This comprehensive framework effectively leverages spatio-temporal correlation to tackle motion artifacts and aliasing artifacts, resulting in improved image details while ensuring computational efficiency.

During the training process, only single-coil raw k-space data was utilized. The provided training dataset included ground truth reference data from 120 subjects, which the Edipo team split in a $90:20:10$ ratio for training, evaluation, and testing, respectively. To enhance the model's robustness, they jointly trained the SAX and LAX data during the training phase. The raw complex data were split into phase and magnitude channels, serving as inputs to both the CRNN module and the DC module. 
For detail refinement in the reconstruction, the intermediate two-channel results were merged into single-channel magnitude data, which was then fed into the refinement module to obtain the final reconstructed image. Training the model employed the Adam optimizer ($\beta1$ = 0.9, $\beta2$ = 0.999, weight decay = 0) and L1 loss, along with SSIM loss, for 50 epochs. The initial learning rate started at $3 \times 10^{-4}$ and gradually reduced to $3 \times 10^{-6}$ with the StepLR scheduler. The GitHub repository for their work is accessible at \url{https://github.com/vios-s/CMRxRECON_Challenge_EDIPO}. 

\subsection{C15 insightdcu}
\begin{figure}[]
\centering
\includegraphics[scale=.5]{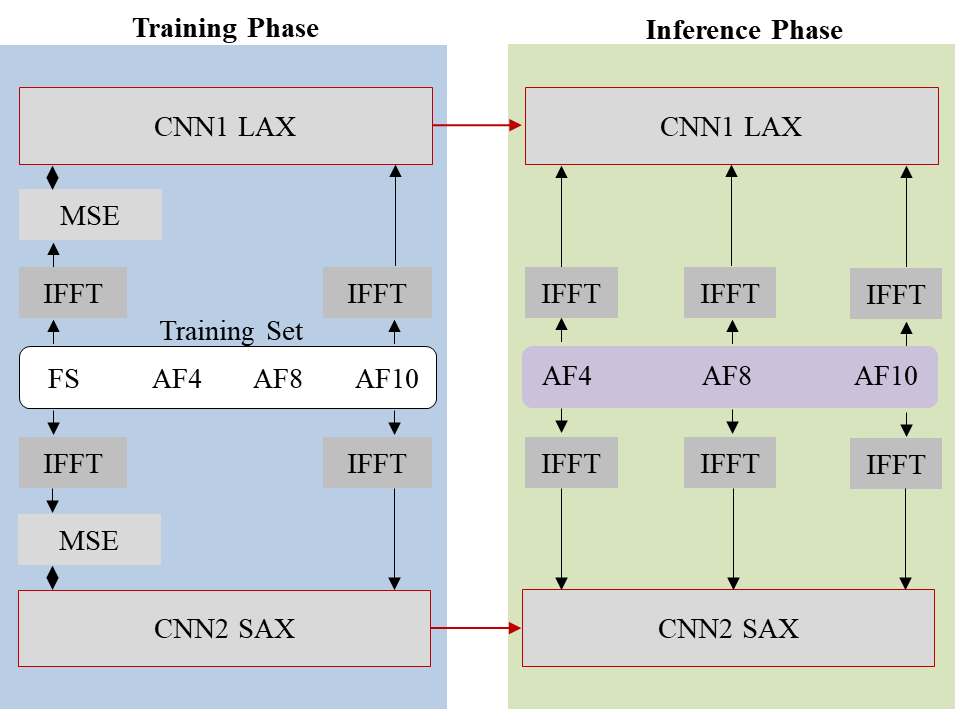}
\caption{Architecture of the proposed double-stream cardiac MRI reconstruction pipeline by team \emph{insightdcu}(C15). Double-stream IFFT and CNN pipeline processes LAX and SAX images separately. The denoising CNN1 and CNN2 are two identical UNET-based models (GNA-UNET). AF4, AF8, and AF10 are abbreviations for acceleration factors 4, 8, and 10, respectively.}
\label{fig:insightdcu}
\end{figure}

The team \emph{insightdcu}(C15) devised a double-stream pipeline to process LAX and SAX data streams independently, as demonstrated in Figure~\ref{fig:insightdcu}. Notably, they focused their denoising CNN training exclusively on $\times$10 undersampled images (AccF10), which exhibit the most prominent aliasing artifacts. Their denoising pipeline employed a UNET-based backbone named GNA-UNET, enriched with Group Normalization (GN) and channel Attention layers on the image domain. This GNA-UNET model adheres to the classical UNET architecture~\citep{ronneberger2015u}, featuring five encoder-decoder blocks and initiating with 64 feature maps in the initial encoder block. Dropout layers were strategically added for regularization. The authors conducted an Ablation Study, revealing that the inclusion of GN layers led to an approximate 2dB PSNR gain when evaluated on a subset of the training set. Additionally, the incorporation of channel attention (Gated Channel Transformation: GCT)~\citep{yang2020gated} layers resulted in a performance gain, albeit of a lower magnitude (0.02dB). 
The encoder block in the GNA-UNET model comprises conv\_block(in\_channels, out\_channels)$\rightarrow$Dropout(p = 0.25)$\rightarrow$MaxPooling2d(2,2) layers. The conv\_block includes conv2d(kernel\_size=3, padding=1)$\rightarrow$GCT$\rightarrow$GN(ng)$\rightarrow$conv2d (kernel\_size=3, padding=1)$\rightarrow$GCT$\rightarrow$GN(ng)$\rightarrow$ReLU. Here, GN(ng) represents a group normalization layer with the hyperparameter ng = 8, determined in the Ablation study. The decoder block mirrors the encoder block in reverse and involves a combination of the transposed convolution layer ConvTranspose2d (kernel\_size=2, stride=2, padding=0)$\rightarrow$conv\_block. The total number of learnable parameters in the GNA-UNET is 124,427,137.

To facilitate rapid experimentation, the team employed a random sampling strategy, selecting 1000 LAX and 1000 SAX images for training from the whole dataset. The loss function utilized was MSE, and AdamW served as the optimizer with a learning rate of 0.001. The model underwent training for 300 epochs to ensure convergence, with no data augmentation applied during this phase. The batch size was set to 2, and the images were resized to a 512$\times$512 input resolution. The images were normalized using a min-max method to the range of [0,1]. 

In conclusion, the pivotal addition of Group Normalization layers in the GNA-UNET model yielded the most substantial performance gain. The GitHub repository for their work is accessible at \url{https://github.com/juliadietlmeier/CMRxRecon_insightdcu}.

\subsection{C17 IADI-IMI}
\begin{figure}[]
\centering
\includegraphics[scale=.36]{ 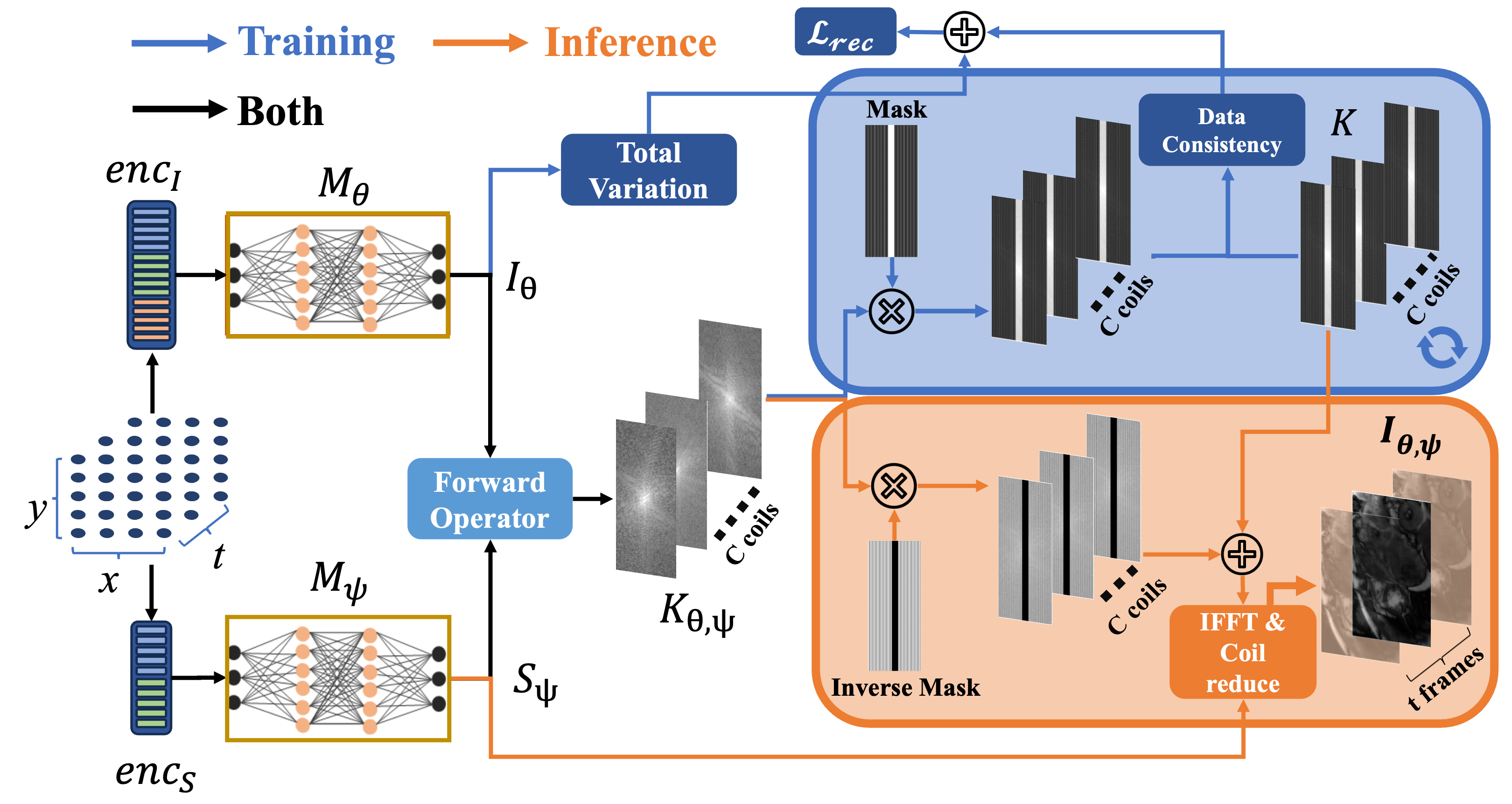}
\caption{The CineJENSE model proposed by the team of \emph{IADI-IMI}(C17) consists of two MLP networks, $M_{\theta}$ and $M_{\psi}$, that receive hash grid encoded 2D+t coordinates as input and predict complex image reconstructions and sensitivity maps, respectively. The forward operator yields coil-expanded k-space predictions that are considered for masked data consistency optimization at training time. At inference time, the inverse mask is used to fill the missing lines of the acquired k-space data, and the final reconstruction is obtained by coil reduction using the estimated sensitivity maps.}
\label{fig:IADIIMI}
\end{figure}

The team \emph{IADI-IMI}(C17) employs an implicit neural representation backbone with multiresolution hash encoding~\citep{muller2022instant} for multi-coil 2D+t cine reconstruction. Their developed CineJENSE, as depicted in Figure~\ref{fig:IADIIMI}, draws inspiration from JSENSE~\citep{ying2007joint} and is an adaptation of IMJENSE~\citep{feng2023imjense}, tailored for dynamic MRI and implicit sensitivity map estimation. The proposed model comprises two shallow MLP networks designed to simultaneously predict a complex-valued intensity reconstruction and a set of complex-valued coil sensitivity maps. Each network is linked to a multiresolution hash grid, mapping spatiotemporal input coordinates to a higher-dimensional encoding optimized for the task. The multiplication of outputs from both MLP networks results in coil-expanded reconstructions, subsequently transformed into multi-coil k-space predictions through Fourier transformation. During training, predictions are masked with the acquisition mask for DC evaluation, while at inference time, predicted k-space lines of the inverse mask are considered to fill the missing lines of the acquired data.

Following an unsupervised instance optimization approach, the model underwent validation on 20 healthy subjects from the training dataset, covering both SAX and LAX cine for all acceleration factors. The proposed 2D+t model processed slices independently while leveraging the temporal information of all cardiac phases. Each undersampled multi-coil 2D+t k-space was scaled by the maximum intensity of its sum-of-squared magnitude reconstruction. Training utilized an Adam optimizer ($\beta1$ = 0.9, $\beta2$ = 0.99) with a learning rate of $1 \times 10^{-2}$ for 200 iterations. To ensure data consistency and denoised intensity reconstructions, the model applied Huber loss ($\delta$=1.0) and total variation regularization, respectively. The GitHub repository for their work is accessible at \url{https://github.com/MDL-UzL/CineJENSE}.

In conclusion, CineJENSE presents a lightweight solution for simultaneous coil sensitivity estimation and cine reconstruction, harnessing spatiotemporal correlations to derive accurate reconstructions from under-sampled k-space acquisitions without the need for fully-sampled reference training data.

\section{Statistical Analysis and Summary of Challenge Outcomes}

\subsection{Overall Outcome}
The challenge has attracted over 285 teams with more than 600 participants. These participants come from 19 countries and regions worldwide. Our official website has received over 15,000 visits, with more than 1,100 users from 38 countries.
We have evaluated over 1,000 submissions during the validation phase (642 submissions for Task 1, and 362 submissions for Task 2), and more than 500 of them have received valid scores on the leaderboard and detailed log files. During the testing phase, we received over 90 Docker images from 22 teams. All the Dockers were successfully run by the organizers.

\subsection{Quantitative and Qualitative Comparisons}

\begin{figure*}[]
\centering
\includegraphics[scale=.35]{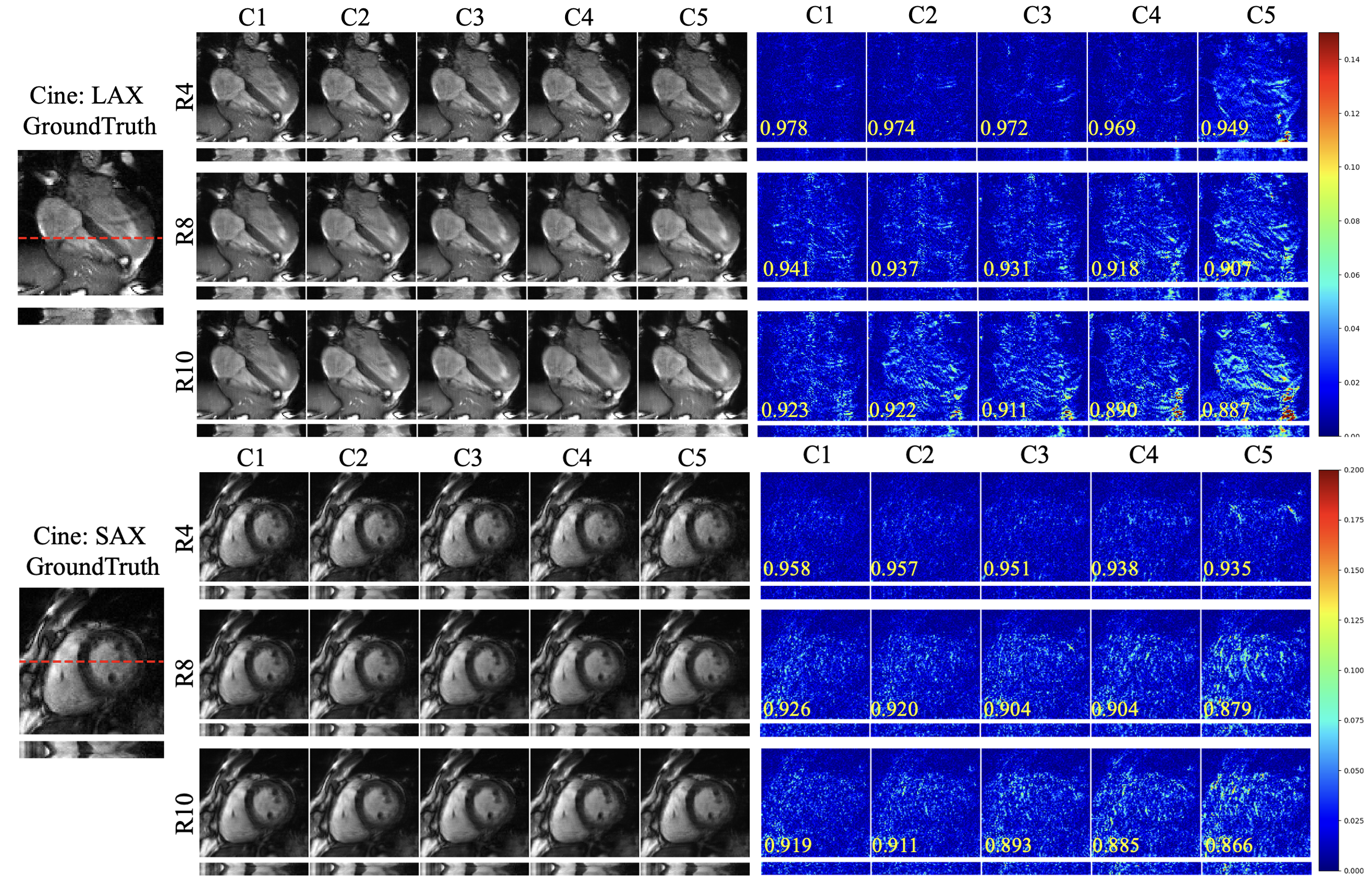}
\label{fig:cine_quali}
\caption{Performances of the top 5 teams in cine task under acceleration factors of 4, 8, and 10. The SSIM of each team is listed in the right bottom corner.}
\end{figure*}

\begin{figure*}[]
\centering
\includegraphics[scale=.3]{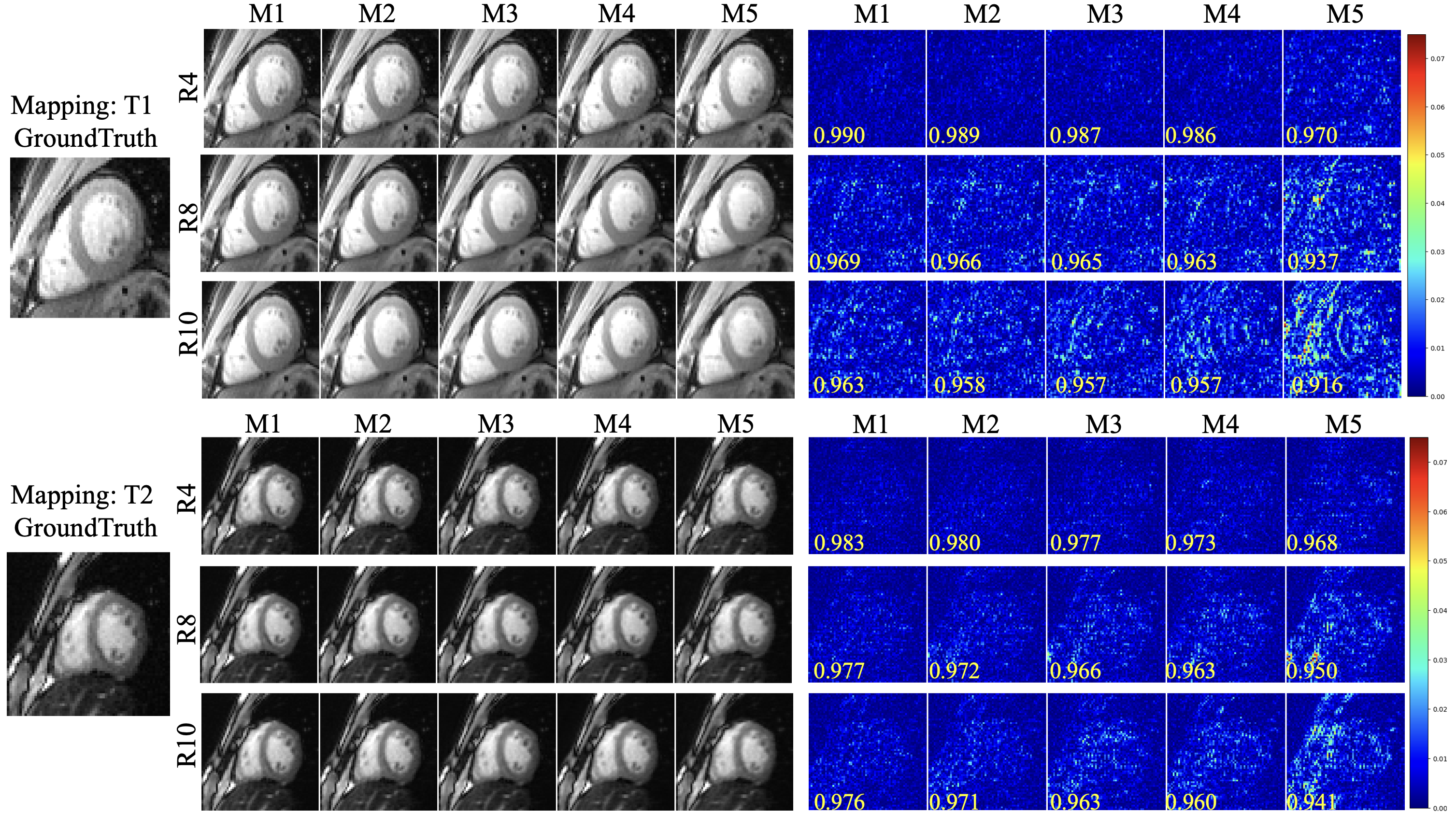}
\caption{Performances of the top 5 teams in mapping task under acceleration factors of 4, 8, and 10. The SSIM of each team is listed in the right bottom corner. }
\label{fig:mapping_quali}
\end{figure*}

\begin{figure*}[]
\centering
\includegraphics[scale=.3]{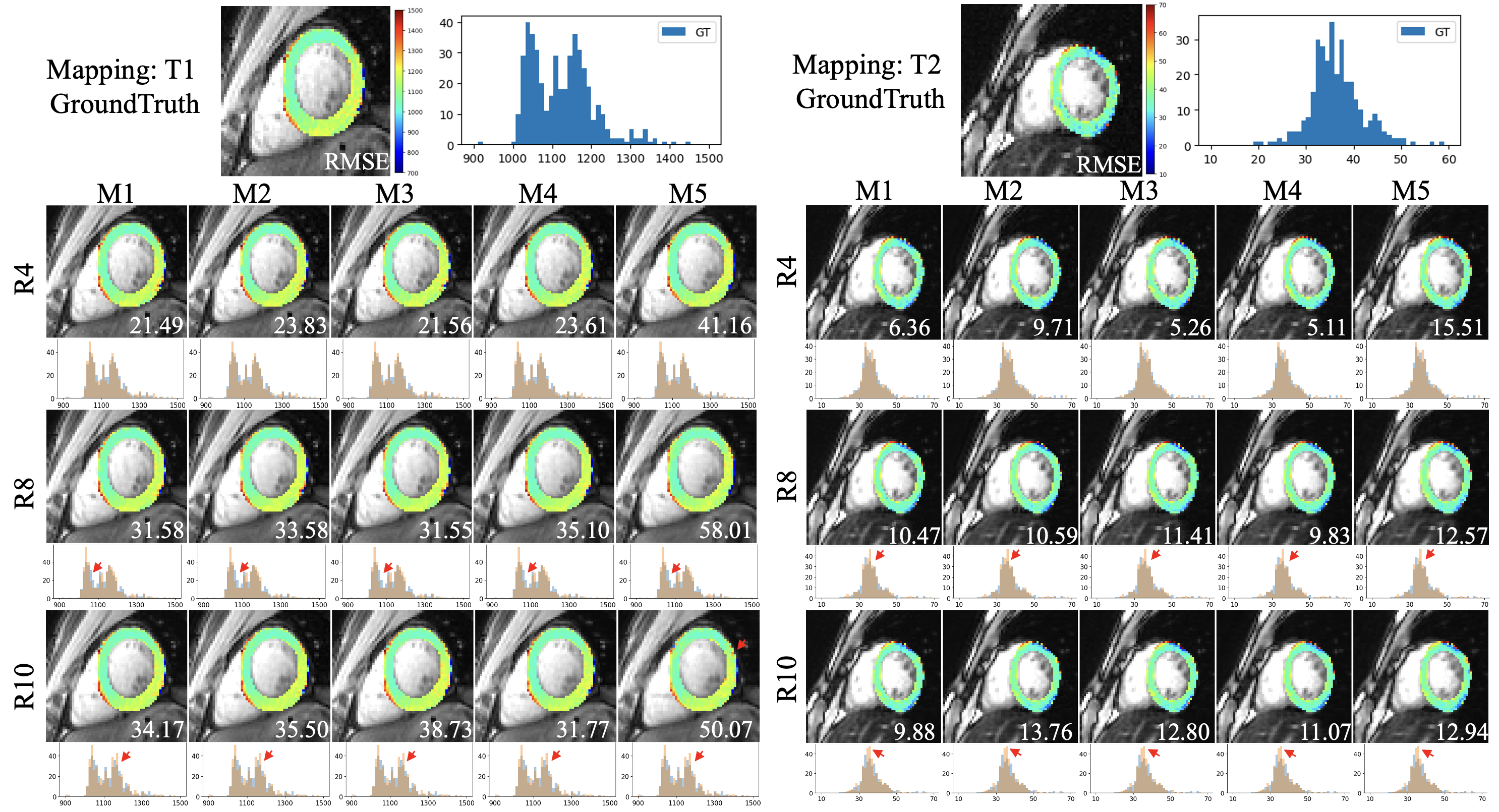}
\caption{The quantification performances of the myocardium are shown in T1/T2 mapping. The histograms of the myocardium are shown. The top 5 teams in mapping tasks under acceleration factors of 4, 8, and 10. The RMSE of each team is listed in the right bottom corner. }
\label{fig:mapping_quali}
\end{figure*}

\begin{table*}[]
\centering
\caption{Evaluation results on  cine of CMRxRecon challenge achieved by participants. The results are reported in the format of mean ± standard deviation.}
\label{tab:cine}
\begin{tabular}{|c|c|c|c|c|}
\hline
{\color[HTML]{330001} Rank} & TeamName                     &  \textbf{SSIM} 
&\textbf{PSNR}  & \textbf{NMSE} \\ \hline
{\color[HTML]{330001} 1}    & hellopipu                    &  0.990±0.002   
&46.873±1.424 & 0.003±0.001  \\ \hline
{\color[HTML]{330001} 2}    & Direct                       &  0.988±0.003   
&46.161±1.381 & 0.004±0.001  \\ \hline
{\color[HTML]{330001} 3}    & clair                        &  0.986±0.003   
&45.221±1.536 & 0.005±0.002  \\ \hline
{\color[HTML]{330001} 4}    & tjubiit                      &  0.984±0.003   
&43.791±1.401 & 0.007±0.002  \\ \hline
{\color[HTML]{330001} 5}    & imr                          &  0.983±0.003   
&43.635±1.456 & 0.007±0.002  \\ \hline
{\color[HTML]{330001} 6}    & Jabber                       &  0.981±0.008   
&35.777±1.308 & 0.048±0.017  \\ \hline
{\color[HTML]{330001} 7}    & SkoICIG                      &  0.969±0.006   
&41.133±1.556 & 0.030±0.006  \\ \hline
{\color[HTML]{330001} 8}    & Mataffine                    &  0.967±0.006   
&39.905±1.371 & 0.014±0.004  \\ \hline
{\color[HTML]{330001} 9}    & OREO&  0.958±0.007   
&37.730±1.371 & 0.024±0.009  \\ \hline
{\color[HTML]{330001} 10}   & feiwang                      &  0.957±0.024   
&39.196±1.439 & 0.019±0.007  \\ \hline
{\color[HTML]{330001} 11}   & Fast2501                     &  0.951±0.015   
&36.443±1.905 & 0.077±0.036  \\ \hline
12                          & Edipo                        &  0.946±0.009   
&35.582±1.313 & 0.037±0.012  \\ \hline
13                          & hkforest                     &  0.945±0.008   
&35.777±1.309 & 0.048±0.017  \\ \hline
14                          & tsinghuacbir                 &  0.923±0.012   
&36.660±1.361 & 0.030±0.010  \\
\hline
15                          & insightdcu                   &  0.921±0.015 
&34.449±1.455 & 0.056±0.022 \\ \hline
16                          & Lyu lab                      &  0.913±0.019 
&32.812±1.847 & 0.093±0.051 \\ \hline
17                          & IADI-IMI                     &  0.911±0.015 
&39.048±1.822 & 0.025±0.016 \\ \hline
18                          & Fzu312lab                    &  0.793±0.098 &25.387±2.458 & 0.588±0.344 \\ \hline
\end{tabular}
\end{table*}

\begin{table*}[]
\centering
\caption{Evaluation results on mapping of CMRxRecon challenge achieved by participants. The results are reported in the format of mean ± standard deviation.}
\label{tab:mapping}
\begin{tabular}{|c|c|c|c|c|c|}
\hline
Rank          & TeamName &  \textbf{SSIM}
&\multicolumn{1}{|c|}{\textbf{PSNR}}  & \multicolumn{1}{|c|}{\textbf{NMSE}} & \multicolumn{1}{|c|}{\textbf{Mapping RMSE}}\\ \hline
1       & hellopipu         &  0.987±0.007      
&45.481±2.705      & 0.004±0.002     &   24.10±1.554        \\ \hline
2       & Direct            &  0.984±0.008      
&44.346±2.600      & 0.004±0.002     &   26.03±1.312        \\ \hline
3 (tie) & clair             &  0.983±0.008      
&43.937±2.527      & 0.005±0.003     &   24.61±1.554     \\ \hline
3 (tie) & dbmapping&  0.983±0.008      
&44.004±2.680      & 0.005±0.003     &   27.35±1.671      \\ \hline
4      & Jabber            &  0.977±0.009      
&41.590±2.333      & 0.008±0.003     &   37.00±2.694      \\ \hline
5     & whitealbum2       &  0.958±0.012      
&38.176±2.401      & 0.033±0.008     &   56.32±7.762       \\ \hline
6     & SkoICIG           &  0.963±0.012      
&39.403±2.436      & 0.030±0.010     &   55.48±8.514       \\ \hline
7     & Fast2501          &  0.934±0.021      
&33.087±2.255      & 0.069±0.032     &   69.10±10.81       \\ \hline
8    & imperial\_cmr     &  0.899±0.025      
&35.628±2.329      & 0.065±0.031     &   66.66±3.293        \\ \hline
9    & IADI-IMI          &  0.812±0.067      
&36.796±2.512      & 0.026±0.013     &   47.39±4.758      \\ \hline
10    & sunnybrook        &  0.771±0.085       &33.690±2.399     & 0.047±0.025    &   60.16±8.104     \\             
\hline
\end{tabular}
\end{table*}

Table~\ref{tab:cine} and Table~\ref{tab:mapping} respectively reported the quantitative evaluation results of cine and mapping reconstruction from different participated teams. Figure~\ref{fig:cine_quali} and Figure~\ref{fig:mapping_quali} show the visualization of the top 5 teams of the two tasks.

The evaluation criteria used in this challenge include PSNR, SSIM, and NMSE. For the final rankings, we considered the higher SSIM value between the multi-coil and single-channel reconstruction results as the final score for ranking. The mean SSIM of images reconstructed from different undersampling rates (R4, R8 and R10) were chosen as the ranking criteria respectively. Quantitative results indicate that the team ``hellopipu" achieved outstanding performance across all metrics. A further RMSE on mapping was evaluated between the myocardium part of the ground truth and reconstructed images. 

For the mapping task, a better SSIM on the original images does not necessarily guarantee improved measurements of the mapping values. For example, ``clair" reached the second lowest RMSE among all the teams but got third place according to both NMSE and SSIM in the mapping task in Table~\ref{tab:mapping}.

\subsection{Consensus on Effective Strategies}
Tables~\ref{tab:CineCharacters} and~\ref{tab:MappingCharacters} respectively outline the key characteristics of the 18 models for cine reconstruction and 11 models for mapping reconstruction. Although we did not specifically encourage teams to perform multi-coil reconstruction, it is worth noting that the majority of participating teams chose to use multi-coil data for reconstruction. The teams used both multi-coil and single-coil data demonstrated improved performance when utilizing multi-coil  reconstruction compared to single-coil approaches, which is within our expectation. In this section, we provide the summary of several effective strategies to cope with CMRxRecon.
These characteristics include the backbone architecture, data standardization, data augmentation strategies, and whether a physical model is employed.

\subsubsection{Loss Function}

\begin{figure*}[]
\centering
\includegraphics[scale=.4]{ 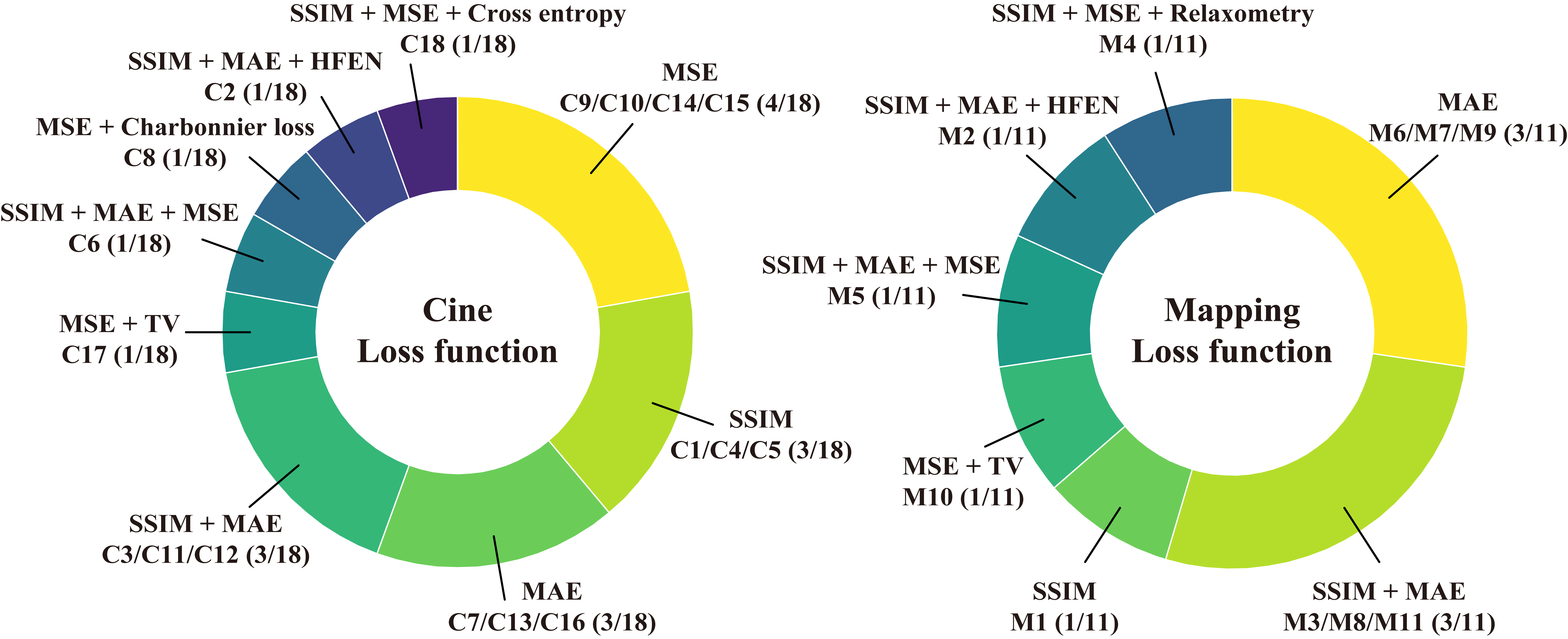}
\caption{Loss function adapted by all ranked teams in the cardiac cine (left) and mapping (right) reconstruction tasks.}
\label{fig:loss}
\end{figure*}

The top 3 performance teams in both cine and mapping tasks all include the SSIM in the loss function, which aligns with the evaluation metrics of the challenge. 

As shown in Figure~\ref{fig:loss}, for cine reconstruction, MSE is the most commonly used loss function among participants, followed closely by SSIM, Mean Absolute Error (MAE), and the composite loss function of MAE+SSIM, which jointly occupy the second position. Additionally, some teams have devised unique composite loss functions, such as MSE+TV, SSIM+MAE+MSE, MSE+Charbonnier, SSIM+MAE+HEFN, and SSIM+MSE+Cross Entropy.

In the case of Mapping reconstruction, MAE and SSIM+MAE are the two most frequently employed loss functions, accounting for half of the participating teams. The remaining five teams have opted for SSIM+MAE+HFEN, SSIM+MSE+Relaxometry, SSIM+MAE+MSE, and MSE+TV, respectively.

\subsubsection{Backbone Architecture}
Several teams utilized external models as backbones, including E2E-VarNet, vSHARP, CAMP-Net, Unet, and Restormer, indicating a reliance on established architectures in Table ~\ref{tab:CineCharacters} and Table ~\ref{tab:MappingCharacters}.
Both the first-place winners of the two tasks utilized the E2E-VarNet \citep{sriram2020end} architecture network as backbones. The fourth-place winner of the cine reconstruction task also use the E2E-VarNet as backbone. In addition, UNet~\citep{falk2019u} architecture is the most common backbone network for both multi-coil and single-channel reconstruction. Convolution neural networks (CNNs) remains the most commonly used backbone model by participating teams, but there are also teams that utilized the Transformer architecture.

\subsubsection{Multi-Scale and Multi-Frame Strategies} 
The majority of teams employed strategies involving multi-scale information fusion and multi-frame training, underscoring the importance of incorporating different scales and temporal information into the modeling process.

\subsubsection{Pre-processings}
Regarding pre-processing, the majority of participating teams opted for normalization of dividing by the maximum value, and a few utilized the z-score normalization method. This step is particularly crucial in the context of challenge, where the original signal intensity in k-space tends to be markedly lower. Data normalization plays a pivotal role in adjusting these intensities to a more suitable scale. This adjustment is essential for the effective operation of activation functions and other components within the neural network. 

\subsubsection{Adherence to Physical Measurements} 
Additionally, the vast majority of participating teams incorporated the physical model, introducing a Data Consistency (DC) \citep{schlemper2017deep} term into the model to ensure consistency between the reconstructed results and the acquired data. In addition to utilizing real-collected k-space data for substitution during the reconstruction process, teams also incorporated various model-based methods into their strategies. These included advanced techniques such as ESPIRiT~\citep{uecker2014espirit} and JSENSE~\citep{ying2007joint} for coil sensitivity estimation, as well as low-rank-based iterative methods for image reconstruction. 

\subsection{Model Complexity Analysis}
\begin{figure*}[]
\centering
    \includegraphics[scale=.4]{ 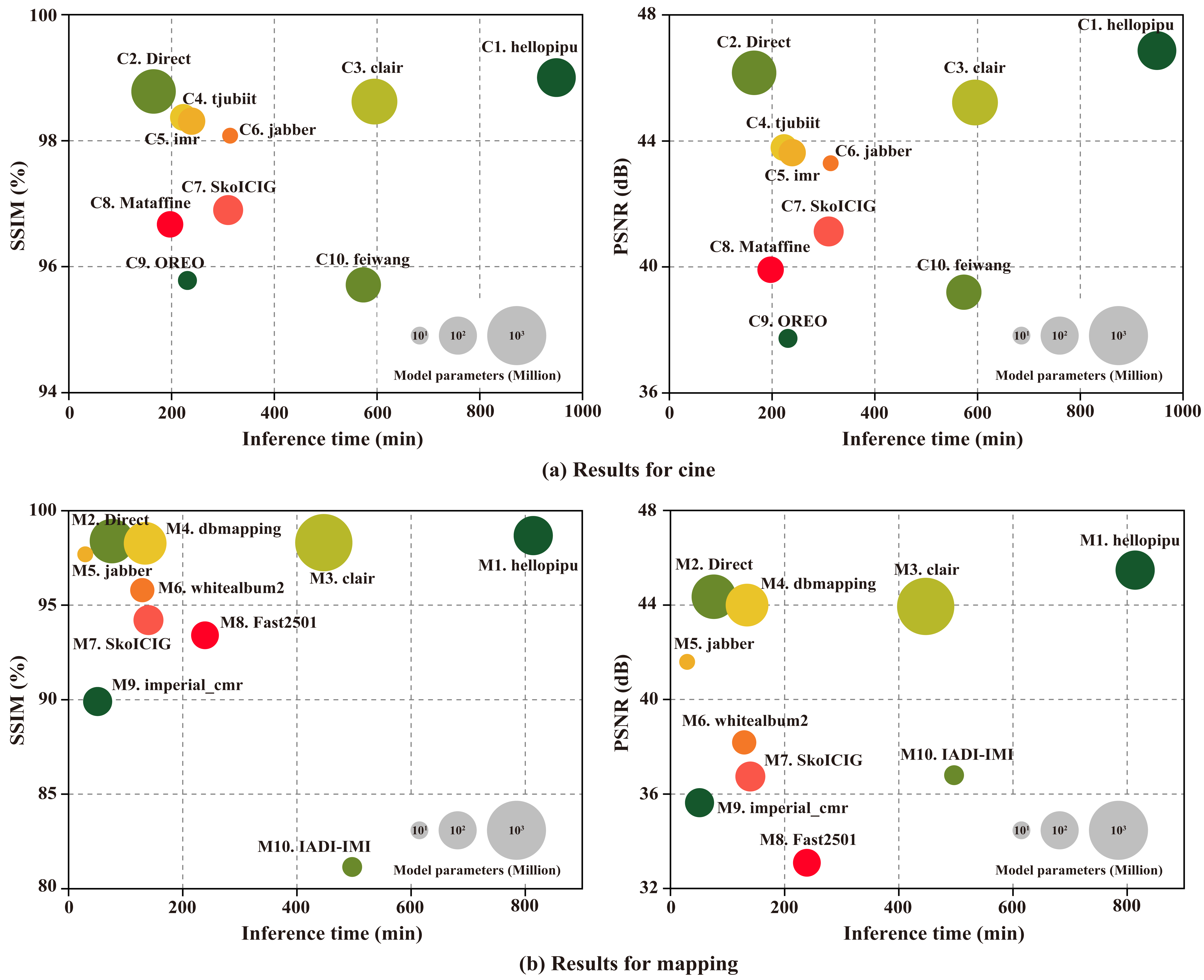}
\caption{Comparison of top 10 teams on the inference times and evaluation metrics. The larger markers indicate more model parameters.}
\label{fig:param}
\end{figure*}

Recently, efficiency has garnered widespread attention in biomedical image processing challenges. The efficiency of cardiac MRI reconstruction methods is particularly crucial in clinical applications. 
Although efficiency does not directly impact rankings, we conducted a supplementary analysis of the complexity of the models from different teams. 
The testing programs submitted by participants were executed on the same Linux workstation, equipped with an Intel(R) Xeon(R) E5-2698 v4 processor (2.20GHz base frequency, 40 cores), 256GB of memory, and one NVIDIA® Tesla V100-DGXS-32GB graphics processors. In competitions such as FLARE'21~\citep{ma2022fast} and ATM’22\citep{zhang2023multi}, the runtime and maximum GPU memory consumption are among the factors considered in the ranking score calculation. 
Tables~\ref{tab:cineinfertime} and~\ref{tab:mappinginfertime}  document the maximum GPU memory consumption and inference time costs for each team in the cine reconstruction and mapping reconstruction tasks, respectively. 
Additionally, in Figure~\ref{fig:param}, we compare metrics based on efficiency. 
C2/M2 demonstrates outstanding overall performance while maintaining a highly level of efficiency. 
In contrast, C1/M1 exhibits longer processing times, possibly due to the iterative algorithm employed, which may increase the inference time costs. 
These findings suggest us that it may be necessary to incorporate model complexity into the evaluation and development of reconstruction methods.

\begin{table*}[]
\footnotesize
\centering
\caption{Statistical analysis for the cine task was performed using the Mann-Whitney U test. This non-parametric test compared the distribution of scores between the highest-scoring model (hellopipu) and each of the other models individually. The resulting P-values are reported for each model comparison.}
\label{tab:ttest-cine}
\resizebox{\textwidth}{!}{
\begin{tabular}{ccccccc}
\hline
Team Name & hellopipu                            & Direct       & clair                        & tjubiit & imr      & Jabber    \\
P-Value   & -                                    & $6.00 \times 10^{-6}$ & $2.89 \times 10^{-13}$                  & $2.35 \times 10^{-17}$ & $5.54 \times 10^{-14}$ & $1.43 \times 10^{-71}$ \\
\hline
Team Name & SkoICIG                              & Mataffine    & OREO and CAFFE MOCHA and HCN & feiwang & Fast2501 & Edipo     \\
P-Value   & $4.01 \times 10^{-30}$ & $4.94 \times 10^{-50}$ & $2.58 \times 10^{-51}$               & $7.32 \times 10^{-85}$ & $7.93 \times 10^{-56}$ & $1.11 \times 10^{-62}$ \\
\hline
Team Name & hkforest                             & tsinghuacbir & insightdcu                   & Lyu lab & IADI-IMI & Fzu312lab \\
P-Value   & $1.21 \times 10^{-71}$ & $2.27 \times 10^{-57}$ & $1.68 \times 10^{-61}$               & $1.17 \times 10^{-57}$ & $7.10 \times 10^{-52}$ & $1.36 \times 10^{-36}$ \\
\hline
\end{tabular}
}
\end{table*}

\begin{table*}[]
\footnotesize
\centering
\caption{Statistical analysis for the mapping task was performed using the Mann-Whitney U test. This non-parametric test compared the distribution of scores between the highest-scoring model (hellopipu) and each of the other models individually. The resulting P-values are reported for each model comparison.}
\label{tab:ttest-mapping}
\resizebox{\textwidth}{!}{
\begin{tabular}{ccccccc}
\hline
Team Name & hellopipu             & Direct                & clair                 & dbmapping-Mapping      & Jabber                & whitealbum2           \\
P-Value   & -                     & $1.01 \times 10^{-7}$ & $3.00 \times 10^{-6}$ & $1.95 \times 10^{-10}$ & $1.60 \times 10^{-35}$ & $4.77 \times 10^{-75}$ \\
\hline
Team Name & SkoICIG               & Fast2501              & imperial\_cmr         & IADI-IMI               & sunnybrook            &                        \\
P-Value   & $2.88 \times 10^{-29}$ & $4.14 \times 10^{-75}$ & $2.95 \times 10^{-147}$ & $4.83 \times 10^{-91}$  & $1.09 \times 10^{-82}$ &                        \\ 
\hline
\end{tabular}
}
\end{table*}

\subsection{Ranking Stability Analysis}
We conducted a Mann-Whitney U test analysis during the final ranking stage. This non-parametric test was performed to compare the distribution of SSIM values between the highest-scoring model (hellopipu) and all other models. As shown in Table~\ref{tab:ttest-cine} and Table~\ref{tab:ttest-mapping}, the SSIM values from all other methods were statistically significantly different from those of hellopipu, with P-value levels below 0.0001.

\begin{table*}[]
\centering
\footnotesize
    \caption{Characteristics of the models from all participated teams in the cardiac cine reconstruction task. Abbreviation: Multi-Coil (MC), Single-Coil (SC), Flip (F), Rotation (R), Shift (S), Data Consistency (DC), FT (Fourier Transform).}
    
    \begin{tabular}{|l|l|l|llll|ll|}
    \hline
    \multirow{2}{*}{\textbf{Team}} & \multirow{2}{*}{\textbf{Backbone}} & \multirow{2}{*}{\textbf{Data standardization}} & \multicolumn{4}{l|}{\textbf{Data augmentation}}  & \multicolumn{2}{l|}{\textbf{Physical model}} \\ \cline{4-9} & & & \multicolumn{1}{l|}{F} & \multicolumn{1}{l|}{R} & \multicolumn{1}{l|}{S} & Others & \multicolumn{1}{l|}{DC}  & Others \\ \hline
    C1. hellopipu & MC, E2E-VarNet \cite{sriram2020end} & Z-score & \multicolumn{1}{l|}{} & \multicolumn{1}{l|}{}  & \multicolumn{1}{l|}{}  & Data balancing                    & \multicolumn{1}{l|}{\checkmark}   & N/A  \\ \hline
    C2. Direct & MC, vSHARP \cite{yiasemis2023vsharp}  & Max  & \multicolumn{1}{l|}{\checkmark}  & \multicolumn{1}{l|}{\checkmark}  & \multicolumn{1}{l|}{} & Multiple undersampling & \multicolumn{1}{l|}{\checkmark}   & ADMM     \\ \hline
    C3. clair & MC, CAMP-Net \cite{zhang2023camp} & Max  & \multicolumn{1}{l|}{}  & \multicolumn{1}{l|}{}  & \multicolumn{1}{l|}{}  & N/A  & \multicolumn{1}{l|}{\checkmark}   & N/A \\ \hline
    C4. tjubiit & MC, E2E-VarNet \cite{sriram2020end}  & Z-score  & \multicolumn{1}{l|}{\checkmark} & \multicolumn{1}{l|}{\checkmark} & \multicolumn{1}{l|}{}  & N/A & \multicolumn{1}{l|}{\checkmark}   & N/A  \\ \hline
    C5. imr & MC, U-Net \cite{falk2019u}& N/A  & \multicolumn{1}{l|}{}  & \multicolumn{1}{l|}{}  & \multicolumn{1}{l|}{}  & N/A  & \multicolumn{1}{l|}{\checkmark}  & N/A  \\ \hline
    C6. jabber & MC, U-Net \cite{falk2019u} & Z-score  & \multicolumn{1}{l|}{\checkmark} & \multicolumn{1}{l|}{}  & \multicolumn{1}{l|}{\checkmark} & N/A  & \multicolumn{1}{l|}{\checkmark}   & N/A  \\ \hline
    C7. SkoICIG  & MC/SC, U-Net \cite{falk2019u}& Z-score & \multicolumn{1}{l|}{}  & \multicolumn{1}{l|}{}  & \multicolumn{1}{l|}{}  & N/A  & \multicolumn{1}{l|}{\checkmark} & N/A \\ \hline
    C8. Mataffine  & MC/SC, U-Net \cite{falk2019u} & N/A  & \multicolumn{1}{l|}{}  & \multicolumn{1}{l|}{}  & \multicolumn{1}{l|}{}  & N/A  & \multicolumn{1}{l|}{}    & N/A      \\ \hline
    C9. OREO & SC, Transformer \cite{vaswani2017transformer}& Max & \multicolumn{1}{l|}{}  & \multicolumn{1}{l|}{}  & \multicolumn{1}{l|}{\checkmark} & N/A  & \multicolumn{1}{l|}{\checkmark}  & N/A \\ \hline
    C10. feiwang & MC, MoDL \cite{aggarwal2018modl} & Max  & \multicolumn{1}{l|}{}  & \multicolumn{1}{l|}{}  & \multicolumn{1}{l|}{}  & N/A & \multicolumn{1}{l|}{\checkmark}   & FISTA    \\ \hline
    C11. Fast2501 & MC/SC, U-Net \cite{falk2019u} & Max & \multicolumn{1}{l|}{\checkmark} & \multicolumn{1}{l|}{}  & \multicolumn{1}{l|}{}  & Multiple undersampling & \multicolumn{1}{l|}{\checkmark}   & ESPIRiT  \\ \hline
    C12. Edipo & SC, CRNN \cite{qin2018convolutional} & N/A  & \multicolumn{1}{l|}{}  & \multicolumn{1}{l|}{}  & \multicolumn{1}{l|}{}  & N/A  & \multicolumn{1}{l|}{\checkmark}   & N/A      \\ \hline
    C13. hkforest & MC/SC, DDPM \cite{Ho2020DDPM} & Max & \multicolumn{1}{l|}{}  & \multicolumn{1}{l|}{}  & \multicolumn{1}{l|}{}  & N/A  & \multicolumn{1}{l|}{\checkmark}   & N/A      \\ \hline
    C14. tsinghuacbir & SC,   CRNN \cite{qin2018convolutional}  & Max & \multicolumn{1}{l|}{\checkmark} & \multicolumn{1}{l|}{}  & \multicolumn{1}{l|}{}  & N/A  & \multicolumn{1}{l|}{\checkmark}   & N/A      \\ \hline
    C15. insightdcu & SC, U-Net \cite{falk2019u} & N/A  & \multicolumn{1}{l|}{}  & \multicolumn{1}{l|}{}  & \multicolumn{1}{l|}{}  & N/A  & \multicolumn{1}{l|}{}    & N/A      \\ \hline
    C16. lyu lab & MC, NAFNet \cite{chu2022NAFNet} & Max & \multicolumn{1}{l|}{}  & \multicolumn{1}{l|}{}  & \multicolumn{1}{l|}{}  & N/A  & \multicolumn{1}{l|}{\checkmark}   & N/A      \\ \hline
    C17. IADI-IMI & MC, INN \cite{Sitzmann2020INN} & N/A  & \multicolumn{1}{l|}{}  & \multicolumn{1}{l|}{}  & \multicolumn{1}{l|}{}  & N/A  & \multicolumn{1}{l|}{\checkmark}   & JSENSE   \\ \hline
    C18. Fzu312lab & SC, U-Net \cite{falk2019u} & Max & \multicolumn{1}{l|}{}  & \multicolumn{1}{l|}{}  & \multicolumn{1}{l|}{}  & N/A  & \multicolumn{1}{l|}{}    & N/A      \\ \hline
    \end{tabular}
    \label{tab:CineCharacters}
\end{table*}

\begin{table*}[]
\footnotesize
\centering
    \caption{Characteristics of the models from all participated teams in the cardiac mapping reconstruction task. Abbreviation: Multi-Coil (MC), Single-Coil (SC), Flip (F), Rotation (R), Shift (S), Data Consistency (DC), FT (Fourier Transform).}
    \begin{tabular}{|l|l|l|llll|ll|}
    \hline
    \multirow{2}{*}{\textbf{Team}} & \multirow{2}{*}{\textbf{Backbone}} & \multirow{2}{*}{\textbf{Data standardization}} & \multicolumn{4}{l|}{\textbf{Data augmentation}} & \multicolumn{2}{l|}{\textbf{Physical model}} \\ \cline{4-9} &  &  & \multicolumn{1}{l|}{F} & \multicolumn{1}{l|}{R} & \multicolumn{1}{l|}{S} & Others  & \multicolumn{1}{l|}{DC}     & Others         \\ \hline
    M1. hellopipu  & MC, E2E-VarNet \cite{sriram2020end} & Z-score & \multicolumn{1}{l|}{} & \multicolumn{1}{l|}{}  & \multicolumn{1}{l|}{}  & Data balancing  & \multicolumn{1}{l|}{\checkmark}      & N/A            \\ \hline
    M2. Direct  & MC, vSHARP \cite{yiasemis2023vsharp} & Max   & \multicolumn{1}{l|}{\checkmark}  & \multicolumn{1}{l|}{\checkmark}  & \multicolumn{1}{l|}{} & Multiple undersampling  & \multicolumn{1}{l|}{\checkmark}      & ADMM  \\ \hline
    M3. clair & MC, CAMP-Net \cite{zhang2023camp}  & Max   & \multicolumn{1}{l|}{}  & \multicolumn{1}{l|}{}  & \multicolumn{1}{l|}{}  & N/A  & \multicolumn{1}{l|}{\checkmark}  & N/A   \\ \hline
    M4. dbmapping  & MC, U-Net \cite{falk2019u} & Min-Max  & \multicolumn{1}{l|}{\checkmark} & \multicolumn{1}{l|}{\checkmark} & \multicolumn{1}{l|}{\checkmark} & Gaussian noise addition & \multicolumn{1}{l|}{\checkmark}  & Relaxometry  \\ \hline
    M5. jabber & MC, U-Net \cite{falk2019u} & Z-score  & \multicolumn{1}{l|}{\checkmark} & \multicolumn{1}{l|}{}  & \multicolumn{1}{l|}{\checkmark} & N/A  & \multicolumn{1}{l|}{\checkmark}      & N/A            \\ \hline
    M6. whitealbum2 & MC/SC, MedNeXt \cite{Roy2023MedNeXt} & Z-score  & \multicolumn{1}{l|}{}  & \multicolumn{1}{l|}{}  & \multicolumn{1}{l|}{}  & N/A  & \multicolumn{1}{l|}{}  & N/A  \\ \hline
    M7. SkoICIG   & MC/SC, U-Net \cite{falk2019u} & Z-score   & \multicolumn{1}{l|}{}  & \multicolumn{1}{l|}{}  & \multicolumn{1}{l|}{}  & N/A   & \multicolumn{1}{l|}{\checkmark}  & N/A  \\ \hline
    M8. Fast2501 & MC/SC, U-Net \cite{falk2019u} & Max   & \multicolumn{1}{l|}{\checkmark} & \multicolumn{1}{l|}{}  & \multicolumn{1}{l|}{}  & Multiple undersampling  & \multicolumn{1}{l|}{\checkmark}  & ESPIRiT \\ \hline
    M9. imperial\_cmr  & MC/SC, MoDL \cite{aggarwal2018modl} & Max  & \multicolumn{1}{l|}{}  & \multicolumn{1}{l|}{}  & \multicolumn{1}{l|}{}  & Multiple undersampling  & \multicolumn{1}{l|}{\checkmark}      & ESPIRiT        \\ \hline
    M10. IADI-IMI  & MC, INN \cite{Sitzmann2020INN} & N/A  & \multicolumn{1}{l|}{}  & \multicolumn{1}{l|}{}  & \multicolumn{1}{l|}{}  & N/A  & \multicolumn{1}{l|}{\checkmark}  & JSENSE  \\ \hline
    M11. sunnybrook  & MC, U-Net \cite{falk2019u} & Z-score  & \multicolumn{1}{l|}{}  & \multicolumn{1}{l|}{}  & \multicolumn{1}{l|}{}  & N/A  & \multicolumn{1}{l|}{\checkmark}  & Low-rank   \\ \hline
    \end{tabular}
    \label{tab:MappingCharacters}
\end{table*}

\begin{table*}[]
\centering
    \caption{Computational consumption and reconstruction performances of top 10 teams in the cardiac cine reconstruction task.}
    \begin{tabular}{|l|l|l|l|l|}
    \hline
    \textbf{Team} & \textbf{CPU memory} & \textbf{GPU memory} & \textbf{Model para.} & \textbf{Inference time} \\ \hline
    C1. hellopipu & 21.97   GB              & 15.78   GB              & 111   M                  & 15h 49min               \\ \hline
    C2. Direct    & 48.22   GB              & 18.3   GB               & 225   M                  & 2h 45min                \\ \hline
    C3. clair     & 145.22   GB             & 2.88   GB               & 264   M                  & 9h 55min                \\ \hline
    C4. tjubiit   & 119.31   GB             & 4.75   GB               & 27 M                     & 3h 43min                \\ \hline
    C5. imr       & 22.18   GB              & 9.18   GB               & 28 M                     & 3h 59min                \\ \hline
    C6. jabber    & 114   GB                & 3.15   GB               & 7 M                      & 5h 14min                \\ \hline
    C7. SkoICIG   & 33.85   GB              & 9.96   GB               & 38 M                     & 5h 10min                \\ \hline
    C8. Mataffine & 23.5   GB               & 6.87   GB               & 25 M                     & 3h 17min                \\ \hline
    C9. OREO      & 19.59   GB              & 3.97   GB               & 10 M                     & 3h 51min                \\ \hline
    C10. feiwang  & 37.98 GB& 4.21 GB& 74 M                     & 9h 43min\\ \hline
    \end{tabular}
    \label{tab:cineinfertime}
\end{table*}

\begin{table*}[]
\centering
    \caption{Computational consumption and reconstruction performances of top 10 teams in the cardiac mapping reconstruction task.}
    \begin{tabular}{|l|l|l|l|l|}
    \hline
    \textbf{Team} & \textbf{CPU memory} & \textbf{GPU memory} & \textbf{Model para.} & \textbf{Inference time} \\ \hline
    M1. hellopipu     & 10.2 GB                 & 18.35   GB              & 121   M                  & 13h 34min               \\ \hline
    M2. Direct        & 23.6 GB                 & 18.08   GB              & 225   M                  & 1h 16min                \\ \hline
    M3. clair         & 50.34 GB                & 5.64   GB               & 1100   M                 & 7h 27min                \\ \hline
    M4. dbmapping     & 29.07 GB                & 14.09   GB              & 181   M                  & 2h 14min                \\ \hline
    M5. jabber        & 48.55 GB& 4.2 GB& 7 M                      & 30 min\\ \hline
    M6. whitealbum2   & 159.52 GB               & 3.46   GB               & 19 M                     & 2h 9min                 \\ \hline
    M7. SkoICIG       & 17.12 GB                & 5.19   GB               & 38 M                     & 2h 20min                \\ \hline
    M8. Fast2501      & 20.78 GB                & 1.94   GB               & 30 M                     & 3h 59min                \\ \hline
    M9. imperial\_cmr & 46.26 GB                & 15.12   GB              & 35 M                     & 51min                   \\ \hline
    M10. IADI-IMI     & 8.56 GB                 & 3.3   GB                & 11 M                     & 8h 17min                \\ \hline
    \end{tabular}
    \label{tab:mappinginfertime}
\end{table*}

\section{Discussion}

\begin{figure}[!t]
\centering
\includegraphics[scale=.45]{ 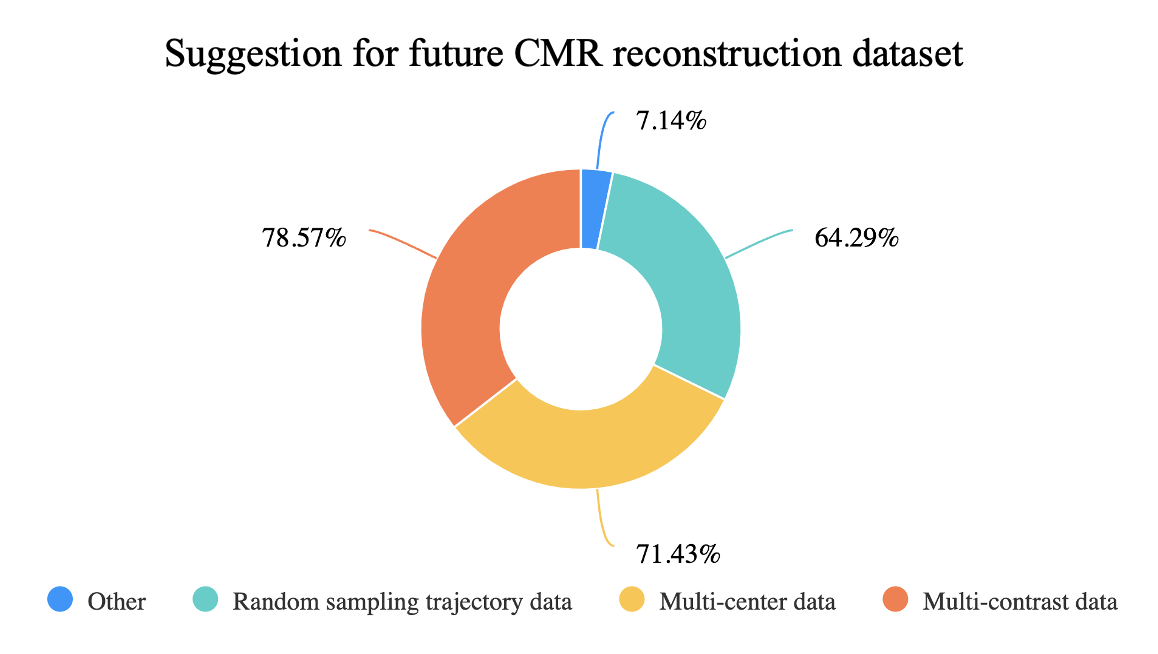}
\caption{Suggestions for future CMR reconstruction dataset from 28 participants in a post-event survey}
\end{figure}

The main goal of CMRxRecon is to provide an open platform for challenge and establish benchmark in the field of cardiac MRI reconstruction in the era of deep learning. The CMRxRecon Challenge has introduced the largest cardiac MRI reconstruction dataset tailored for deep learning applications to date. However, cardiac MRI reconstruction remains an open challenge for the research community. The analysis of the winning solution from `hellopipu' reveals that while significance has been made with PSNR, SSIM, and NMSE scored 46.873, 0.990, 0.003, and 45.481, 0.987, 0.004 for cine and mapping reconstruction, respectively, the exploration into the nuanced recovery of details within the reconstructed images remains challenging, especially for higher undersampling factors. To better assess the application potential of reconstruction models, it will be necessary to  introduce some patient data for testing. Specifically, based on the T1 and T2 values calculated from the reconstructed images, `clair', who ranked the third place, has outperformed `Direct' in terms of RMSE compared with the ground-truth mapping values~\ref{tab:mapping}.

As shown in Table~\ref{tab:cineinfertime} and Table~\ref{tab:mappinginfertime}, the inference time of hellopipu's model is about 7 minutes and 6 minutes per case for cine and mapping tasks, respectively, since hellopipu employs models with 111M parameters for the cine reconstruction and 121M parameters for the mapping reconstruction. While not the highest parameter count reported, the model's complexity could contribute to longer processing times. However, the inference time does not solely depend on the model size, as demonstrated by other teams with larger models achieving faster inference times. The CPU memory usage by `hellopipu' is relatively moderate in cine reconstruction and significantly lower in mapping reconstruction compared to the highest CPU memory usage reported. However, GPU memory usage is among the highest for both tasks. Since they have incorporated the large foundation model as the Prompt Block, how to use better synchronization, communication overhead or optimization methods is very crucial. Moreover, we just use one GPU in the testing stage. Ensuring efficient inference when restricted to using just one GPU centers around maximizing the computational efficiency and throughput of the available hardware. This process involves a combination of optimizing the model itself and leveraging the specific capabilities and features of the GPU to full effect. Therefore, the trade-off between the reconstruction performance and the model's extrapolation time remains a promising research direction.

In a post-event survey, suggestions for future improvements to the CMR reconstruction challenge were received from 28 participants. A total of 78.58\% of teams expressed a desire for k-space data with more contrast in future challenges. Additionally, 71.43\% of teams recommended the inclusion of multi-center cardiac data to enhance the diversity of the dataset. Furthermore, 64.29\% of teams hoped for the availability of random sampling trajectory data, contributing to a more comprehensive coverage of cardiac images under different scenarios. The remaining 7.14\% proposed other content. These suggestions will contribute to optimizing future CMR challenges as well as public datasets to better align with the needs from the participants. Our next step involves expanding in the following areas as we continue to organize several CMR reconstruction challenges in the future:

\noindent \textbf{Providing diverse sampling trajectories.} Instead of applying uniform sampling along the time dimension, better strategy should follow random sampling, hence maximizing the information along the time dimension. Such a refined strategy not only enhances the efficiency of image acquisition but also contributes significantly to the integration of CMR imaging into clinical workflows.

\noindent \textbf{Expanding clinical generalization through multi-center and disease-specific Data.} Our current dataset~\citep{wang2023cmrxrecon} predominantly consists of CMR data from healthy volunteers, obtained using equipment from a single vendor in a single-center setting. Recognizing the limitations this imposes on the generalizability of data-driven models, we suppose to include a diverse range of pathological conditions and data from multiple vendors and centers. This comprehensive inclusion will pave the way for the development of robust models capable of accommodating the variability inherent in clinical environments, thereby enhancing the reliability and applicability of CMR across different patient populations and diagnostic contexts.

\noindent \textbf{Finding advanced evaluations for performance benchmarking.} We leveraged the SSIM as our evaluation for benchmarking in the current challenge, which enables a precise comparison between the reconstructed images and fully-sampled images. However, we acknowledge the necessity of a more encompassing evaluation framework that goes beyond mere image quality, such as inference time and the generation of further parametric maps. Inference time, indicative of the speed at which the models reconstruct images, is paramount in clinical settings where timely diagnostics are critical. Moreover, the development and analysis of parametric maps extend our capabilities beyond anatomical imaging, offering insights into the functional and tissue characteristics of the myocardium.

\noindent \textbf{Trustworthy reconstruction on multi-contrast CMR imaging.} We aim to obtain data with more contrasts to achieve reliable and accurate reconstructions. The complexity and diversity of CMR scans in real-world applications, involving various contrasts, sampling trajectories, scan orientations, equipment vendors, and disease types, present a great challenge for existing AI-based reconstruction methods, which are usually developed for only one or a few specific scanning settings. In practice, there are often inevitable domain mismatches between the training data and target data, due to the diversities listed above \citep{ouyang2019generalising,knoll2019assessment,yang2023physics}. Therefore, building and validating universal and robust reconstruction models for handling these diversities remains a critical technical challenge for multi-parametric CMR imaging \citep{ouyang2019generalising,liu2021universal,wang2023pisf,tanzer2023t1}. To accomplish this, people may leverage a universal pre-trained reconstruction model to handle the heterogeneity and intricacies of multi-contrast imaging, ensuring high fidelity and trustworthiness in reconstructed results. 

\section{Conclusion}
The CMRxRecon challenge offers a benchmark dataset comprising multi-contrast, multi-view, and multi-coil raw k-space data with manually annotated labels for cardiac anatomical structures. This dataset enables the research community to actively contribute to the development of deep learning-based cardiac MRI reconstruction algorithms. Our paper provides a comprehensive overview of the challenge design, presents a summary of the submitted results, reviews the employed methods, and offers an in-depth discussion that aims to inspire future advancements in cardiac MRI reconstruction models. The summary highlights effective strategies observed in cardiac MRI reconstruction, including backbone architecture, loss function, pre-processing techniques, physical modeling, and model complexity, providing valuable insights for further advancements in this field.

\section*{Acknowledgments}
This work was supported in part by the National Natural Science Foundation of China under Grants 62371413, 62331021, and 62122064, in part by Yantai Basic Research Key Project 2023JCYJ041, in part by the Youth Innovation Science and Technology Support Program of Shandong Provincial under Grant 2023KJ239, in part by the Natural Science Foundation of Fujian Province of China under Grant 2023J02005, in part by the President Fund of Xiamen University under Grant 20720220063, in part by the EPSRC, UK Grants (TrustMRI: EP/X039277/1), in part by the UKRI Centre for Doctoral Training in AI for Healthcare, Imperial College London under Grant EP/S023283/1,  in part by the ERC IMI (101005122), the H2020 (952172), the MRC (MC/PC/21013), the Royal Society (IEC$\backslash$NSFC$\backslash$211235), the NVIDIA Academic Hardware Grant Program, the SABER project supported by Boehringer Ingelheim Ltd, and the UKRI Future Leaders Fellowship (MR/V023799/1), in part by the National Institutes of Health (NIH) grant 7R01HL148788-03, the Royal Academy of Engineering and the Research Chairs and Senior Research Fellowships scheme (grant RCSRF1819$\backslash$8$\backslash$25), and the UK’s Engineering and Physical Sciences Research Council (EPSRC) support via grant EP/X017680/1, in part by the China Scholarship Council under grant 202306310177.
The computations in this research were performed using the CFFF platform of Fudan University. 

\section*{Author contributions}
C.W: Project administration, Conceptualization, Methodology, Validation, Data curation, Writing, review and editing; J.L, C.Q: Conceptualization, Methodology, Validation, Formal analysis, Data curation, Writing, review and editing ; S.W: Conceptualization, Validation; F.W: Data curation, Software, Formal analysis, Writing original draft; Y.L: Data curation, Validation, Critical Evaluation; Z.W: Software, Formal analysis, Writing original draft; K.G: Software, Methodology; C.O: Critical Evaluation, Review and editing; M.T: Formal analysis, Methodology, Writing original draft; M.L, L.S, M.S, Q.L, Z.Z: Software, Validation, Critical Evaluation; Z.S, S.H: Clinical evaluation, Data curation, Validation; H.L, Z.C: Critical Evaluation, Review and editing; Z.X: Data curation, Software, Coordination; Y.Z, Y.C, W.C: Critical Evaluation, Review and editing; W.B, X.Z, J.Q, L.W, G.Y, X.Q, H.W: Supervision, Conceptualization, Review; B.X, D.M, G.Y, J.T, L.Z, W.C, Y.P, X.L, A.R, D.D, Q.D, K.Y, Y.X, Y.D, J.D, C.G.C, Z.A.H, N.V were participants of the CMRxRecon challenge, and provided their results for evaluation and the description of their algorithms. The final manuscript was approved by all authors.

\section*{Declaration of Competing Interest 
}
The authors declare that they have no competing financial interests or personal relationships that could be appeared to influence the work reported in this paper. 

\bibliographystyle{model2-names.bst}\biboptions{authoryear}
\bibliography{refs}



\end{document}